\documentclass[12pt,preprint]{aastex}
\usepackage{lscape}
\usepackage{graphicx}

\newcommand{\rsun}{{\rm \ R_\odot}}

\newcommand\teff{T$_{\rm eff}$}
\newcommand\logg{log $g$}

\def\moaspaces{{MOA-2006-BLG-99S}\ }
\def\moas{{MOA-2006-BLG-99S}}
\def\moaspace{{MOA-2006-BLG-99}\ }

\def\oglespace{{OGLE-2006-BLG-265S}\ }
\def\ogle{{OGLE-2006-BLG-265S}}

\def\max{{\rm max}}

\def\deg{\ensuremath{^\circ}}
\begin{document}
\title{A High-Resolution Spectrum of the Highly Magnified 
 Bulge G-Dwarf MOA-2006-BLG-099S\altaffilmark{1}}

\author{
Jennifer~A.~Johnson\altaffilmark{2},
B.~Scott Gaudi\altaffilmark{2},
Takahiro Sumi\altaffilmark{3},
Ian A. Bond\altaffilmark{4}
and
Andrew~Gould\altaffilmark{2}
}
\altaffiltext{1}{This paper includes data gathered with the 6.5 meter Magellan Telescopes located at Las Campanas Observatory, Chile.}
\altaffiltext{2}
{Department of Astronomy, Ohio State University,
140 W.\ 18th Ave., Columbus, OH 43210, USA; 
jaj,gaudi,gould@astronomy.ohio-state.edu}
\altaffiltext{3}
{Solar-Terrestrial Environment Laboratory
Nagoya University, Nagoya, Japan;sumi@stelab.nagoya-u.ac.jp }
\altaffiltext{4}
{Institute for Information and Mathematical Sciences, Massey University,
Auckland, New Zealand;i.a.bond@massey.ac.nz}

\begin{abstract}

We analyze a high-resolution spectrum of a microlensed G-dwarf in the
Galactic bulge, acquired when the star was magnified by a factor of
110. We measure a spectroscopic temperature, derived from the wings of the
Balmer lines, that is the same as the photometric temperature, derived using
the color determined by standard microlensing techniques.
We measure [Fe/H]=$0.36 \pm 0.18$, which places this star at the
upper end of the Bulge giant metallicity distribution.  In particular,
this star is more metal-rich than any bulge M giant with high-resolution
abundances.  We find that the abundance ratios of alpha and iron-peak
elements are similar to those of Bulge giants with the same
metallicity.  For the first time, we measure the abundances of K and Zn
for a star in the Bulge. 
The [K/Mg] ratio is
similar to the value measured in the halo and the disk, suggesting
that K production closely tracks $\alpha$ production. The [Cu/Fe]
and [Zn/Fe] ratios support the theory that those elements are produced in
Type II SNe, 
rather than Type Ia SNe. We also measured the first C and
N abundances in the Bulge that have not been affected by first
dredge-up. The [C/Fe] and [N/Fe] ratios are close to solar, in
agreement with the hypothesis that giants experience only canonical mixing.

\end{abstract}


\section{Introduction
\label{sec:intro}}

The Galactic Bulge underwent an intense burst of star
formation early in the formation of the Galaxy, leading to a very
different stellar population and chemical evolution history than found
in the Milky Way disk or halo
\citep[e.g.,][]{ortolani:95,zoccali:03,mcwilliam:94}. In particular,
massive stars
may dominate the pollution at almost all metallicities, leading to unique
abundance patterns
\citep[e.g.,][]{lecureur:07}. Similar events are thought to mark the formation
of other galactic spheroids, making the Bulge stellar population a
template for interpreting extragalactic observations. As a result of
its unique formation history in the Galaxy, the Bulge has been the
subject of intensive study.

The detection of RR Lyrae stars \citep{baade:46} first indicated that the Bulge
contained old stars. With deeper photometry, the main sequence turnoff
(MSTO) of the
Bulge was detected. \citet{terndrup:88} found a mean age of 11-14 Gyr for
stars in Baade's Window, with a negligible fraction of stars with ages
$<$ 5 Gyr. Photometry reaching more than 2 magnitudes below the MSTO with {\it Hubble Space Telescope} confirmed the generally old
nature of the Bulge \citep{ortolani:95,holtzman:98}, although
\citet{feltzing:00} included a reminder that a young metal-rich
population has similar MSTO colors and luminosities to an
older, more metal-poor population.
Therefore, deriving ages reliably from photometry of the MSTO requires
adequate knowledge of the Bulge metallicity distribution
function (MDF), specifically for the MSTO stars. However,
because of the faintness of those stars, the measurement of the Bulge MDF 
has historically relied on giants.

Since the discovery of both M giants and RR Lyr stars, it has been
known that the Bulge giants span a wide range in
metallicity. Low-dispersion spectra provided the first quantitative
measure of the MDF \citep{whitford:83,rich:88}.  \citet{sadler:96}
measured indices from low-dispersion spectra of 268 K giants (both
red clump stars and first ascent giants) 
to derive a mean metallicity\footnote{We adopt the usual spectroscopic notation that [A/B] 
$\equiv$ {\rm log}$_{\rm 10}$(N$_{\rm A}$/N$_{\rm B}$)$_{\star}$ 
-- {\rm log}$_{\rm 10}$(N$_{\rm A}$/N$_{\rm B}$)$_{\odot}$}) 
$\langle{\rm [Fe/H]}\rangle =-0.11$, with a dispersion of 0.46 dex. 
Recalibration by \citet{fulbright:06} based on high-resolution
spectra of 15 stars in common with the \citet{sadler:96} sample reduced
the mean metallicity to $-0.22$.  \citet{ramirez:00} measured
$\langle{\rm [Fe/H]}\rangle =-0.21$ 
from low-dispersion near-infrared spectra 
for 72 M giants in the inner Bulge.
The good agreement between \citet{ramirez:00} result and the 
recalibrated \citet{sadler:96} result is somewhat surprising. 
At the bright end of the giant branch, 
only metal-rich first ascent giants become M giants. However, both
K giants and M giants become red clump stars, and lower luminosity
metal-rich giants are K stars as well. Therefore, the
inclusion of red clump stars  and fainter giants 
in the \citet{sadler:96}
sample make the biases in their sample more similar to those of the
\citet{ramirez:00} study.

\citet{zoccali:03} measured both the MDF of the bulge and the
age of the stars using deep photometry of the Bulge in the optical
and near-infrared wavelengths. They
constructed the M$_{K}$ and (V-K)$_0$ color magnitude diagram for a
low-reddening window at ($l,b$)=(0.277, $-$6.167). Because the slope
of the red giant branch (RGB) in these colors depends on the metallicity, RGB
stars with different metallicities fall on different parts of the 
color-magnitude diagram. They could therefore 
use their photometry to derive the MDF of 503 giants by
comparing the positions of bright (M$_K < -4.5$) RGB stars with
globular cluster fiducials of known metallicity. 
After correcting for small biases in their MDF caused by their magnitude
cutoff, they find an MDF with a
peak at [M/H]$=-0.1$, a sharp cutoff at [M/H]$=-0.2$ and few stars with
[M/H]$<-1$. Adopting the metallicities derived from the giants for the
MSTO dwarfs, they estimated that the
Bulge is coeval with the halo and argued that the lack of stars above
the prominent MSTO of the Bulge ruled out a significantly younger
population.

These measurements of the Bulge giant MDF can be improved by metallicity
measurements from high-dispersion measurements of many stars in 
several fields throughout the Bulge. Recently, \citet{lecureur:08} and 
\citet{zoccali:08}  have obtained a total of $\sim 1000$ K giant spectra at
(R$\sim$20,000) in 4 Bulge windows. They confirm the metal-rich
nature of the Bulge.

With such a metal-rich population having been reached so quickly after star
formation began, the Bulge is expected to have a distinct chemical
evolution compared to other Galactic populations, such as the halo or
the disk because, for example, the contributions of longer-lived polluters,
such as Type Ia supernovae (SNe) or low-mass AGB stars, should be small.
Indeed, \citet{mcwilliam:94} measured high [$\alpha$/Fe] ratios in giants, in
particular, high [Mg/Fe] for [Fe/H] values up to solar, arguing for little Type
Ia SN contribution of Fe compared to the thick or thin
disks. \citet{fulbright:07} confirmed the overall enhancement in [Mg/Fe]
and strengthened the conclusion of \citet{mcwilliam:94} that the other
$\alpha$ abundance ratios do not track [Mg/Fe] exactly. [O/Fe],
[Si/Fe], [Ca/Fe] and [Ti/Fe] begin to decrease around [Fe/H]=0, while
[Mg/Fe] does so at supersolar [Fe/H].  \citet{fulbright:07} suggested that
metallicity-dependent Type II SN yields could explain the different
behaviors of the $\alpha$ elements. \citet{mcwilliam:07} explained the
low [O/Mg] ratios in metal-rich Bulge giants through a different 
metallicity-dependent mechanism: Wolf-Rayet winds
leading to less effective O production in metal-rich massive stars.
By [Fe/H]$\sim$0.2, all the [$\alpha$/Fe] ratios have
begun to decline, probably indicating the introduction of large
amounts of Fe from Type Ia SNe \citep[e.g.,][]{cunha:06}.

Several studies have looked at the abundances of the light elements Na and
Al in the Bulge, two elements whose production should depend on the 
metallicity of the massive stars that exploded as Type II SNe.
\citet{cunha:06} measured Na in 7 K and M giants and found the
predicted increase in [Na/Fe] and [Na/O] 
in the most metal-rich stars. \citet{lecureur:07} found supersolar 
[Al/Fe] at all metallicities and, for [Fe/H] $>$0, enhanced [Na/Fe]
compared to the ratios in disk stars. Interestingly, at higher
metallicities, the scatter in [Al/Fe] and [Na/Fe] increased and became
larger than could be explained by observational errors. Interpreting
the Na and Al abundances as the result of Type II SN production may
be problematic. The surface abundances of Al and Na have
been shown to be increased by large amounts of internal mixing in 
metal-poor globular cluster stars \citep[e.g.][]{shetrone:96}, 
where the products of  proton-capture
reactions deep inside the star are mixed up to the surface, leading
to enhancements in these two elements. However, \citet{lecureur:07}
argued that the C and N abundances in the giants they studied were 
consistent with only mild mixing and, therefore, that the high Na and Al had
to be due to the overall chemical evolution of the Bulge. \citet{cunha:06}
also found evidence for mild mixing in giants, 
affecting C and N, but not O, Na, or Al.

Finally, there is little information on the neutron-capture elements in
the Bulge. The absorption lines for these elements are concentrated in the
blue part of the optical spectrum, where the crowding from Fe, CN, and other
lines is severe. Near-IR spectra have essentially no lines of 
these elements. While there
are a few lines of Ba in the red, these lines in metal-rich giants are
so saturated that reliable measurements are very difficult. As a result,
only the neutron-capture element Eu has published results so far. 
\citet{mcwilliam:94} found [Eu/Fe]$>$0 in Bulge giants, which is likely
because of the production of Eu in the r-process. Measuring additional 
neutron-capture element abundances would test this idea, because the
r-process is better at making some elements (e.g., Eu) than others (e.g., Ba).

Our knowledge of the metallicity and abundance ratios of
Bulge stars has generally been confined to the bright giants, which are usually the
only ones accessible to high-resolution observations.  But studying
the dwarfs has several advantages.  Their abundances of the elements such as C and N
are unaffected by dredge-up processes on the giant branch. We can
measure elements, such as S and Zn, that not measured in giants,
because the hotter temperatures of the dwarfs decrease the strength of
CN and increase the strength of certain atomic lines. In addition, it
is critical to know the metallicity of stars at the MSTO
to accurately measure the ages from their color and luminosity. 
Finally, individual ages can be assigned to dwarf stars
near the turnoff.

The advent of large surveys to identify and follow-up microlensing
events, such as the 
Microlensing Observations in Astrophsycis (MOA) 
collaboration\footnote{http://www.massey.ac.nz/$\sim$iabond/alert/alert.html},
the Optical Gravitational Lens 
Experiment\footnote{http://www.astrouw.edu.pl/$\sim$~ogle/ogle3/ews/ews.html}
(OGLE), the Microlensing Follow Up 
Network\footnote{http://www.astronomy.ohio-state.edu/$\sim$microfun/}
($\mu$FUN) and the Probing Lensing Anomalies 
Network\footnote{http://planet.iap.fr/} (PLANET), provides an
opportunity to study otherwise unobservable Bulge dwarfs.  During
high-magnification microlensing events, it is possible to obtain
high-resolution, high signal-to-noise ratio spectra of faint stars
with a huge savings in observing time: a factor $A\times 10^{0.4\Delta
m}$ where $A$ is the magnification and $\Delta m$ is the number of
magnitudes below sky of the unmagnified star.  In \citet{johnson:07} ,
we reported the detailed abundances for a highly-magnified Bulge
dwarf, \ogle, which from a 15 minute
exposure at magnification $A\sim 135$, was shown to be one of the most metal-rich
stars ever observed.  It also provided the first measurements of S and
Cu in the Bulge.  Here we present a high-resolution spectrum of the
Bulge G-dwarf MOA-2006-BLG-99S, taken at magnification $A=110$.

Finally, we respond the challenge: ``Ask not what microlensing can
do for stellar spectroscopy -- ask what stellar spectroscopy can
do for microlensing.'' There is one important way
that the spectroscopic study of bulge dwarfs can benefit microlensing.
Whenever a source approaches or transits a ``caustic''
(line of infinite magnification) caused by the lens, one can measure
$\rho$, the ratio of the angular source radius to angular Einstein radius
$\rho=\theta_*/\theta_{\rm E}$, from the microlens lightcurve.
Then $\theta_*$ is inferred from the dereddened color and magnitude
of the source to yield $\theta_{\rm E}=\theta_*/\rho$,
which in turn provides important constraints on the lens properties.
Because spectroscopy is not normally available for these microlensed
sources, the dereddened color and magnitude are estimated by
comparing the source position on an instrumental color-magnitude diagram
with that of the clump and then assuming that the Bulge clump is similar to
the local clump as measured by {\it Hipparcos} \citep{ob03262}.  This
procedure undoubtedly suffers some statistical errors and could
suffer systematic errors as well.  For example, the Bulge clump
may have a different color from the local one.  High-resolution
spectra of an ensemble of microlensed bulge sources will test
this procedure for both statistical and systematic errors.
For OGLE-2006-BLG-265S, the standard microlensing procedure yielded
$(V-I)_0=0.63\pm 0.05$, whereas \citet{johnson:07} obtained
$(V-I)_0=0.705\pm 0.04$ from high-resolution spectroscopy.  This
difference hints at a possible discrepancy, but only by repeating
this procedure on a number of dwarfs can this be confirmed.

\section{Observations
\label{sec:data}}

\moaspace was alerted as a probable microlensing event 
toward the Galactic Bulge (J2000 RA = 17:54:10.99, Dec = $-$35:13:38.0;
$l=-4.48$, $b=-4.78$) by MOA on 22 July 2006.  On 23 July, MOA
issued a further alert that this would be a high-magnification
event, with $A>100$.  Intensive photometric observations were then
carried by several collaborations, including $\mu$FUN,
primarily with the aim of searching for planets 
\citep{mao91,griestsafi,ob05071,ob05169}.
Results of that
search will be presented elsewhere.  The event actually peaked
on 23 July (HJD 2453940.349) at $A_\max\sim 350$.  One of us 
(BSG) happened to be at the Clay 6.5-m Telescope at the
Magellan Observatory when he received the flurry of 
$\mu$FUN emails describing this event.  He then interrupted his
normal program to obtain a 20-minute exposure of this event at the
beginning of the night, just after peak, when the magnification was
$A=215$.  Unfortunately, the atmospheric transparency was poor, and the
observation had to be interrupted after 1089 seconds when the cloud cover
became too thick.  Conditions cleared several hours later, and he
obtained two 20-minute exposures centered at UT 02:09:36 and UT 02:30:34 
24 July, when
the magnification was $A=113$ and $A=107$.  We base our results 
on these higher quality spectra.

The observations of MOA-2006-BLG-99 were made
using the Magellan Inamori Kyocera Echelle (MIKE) double spectrograph
\citep{bernstein:03} mounted on the Clay telescope on Las Campanas.
with seeing of 0.7--1.0 arcsec. We used the 1.0 arcsec slit, which
produces R$\sim$25,000 on the blue side and R$\sim$19,000 on the red
side.

\section{Data Reduction}

The data were reduced using the MIKE Python data reduction pipeline 
(D. Kelson, 2007,
private communication), with the exception of the bluest orders containing
the CH and CN lines, which were reduced using the IRAF\footnote{IRAF is distributed by the National
Optical Astronomy Observatories, which are operated by the Association of
Universities for Research in Astronomy, Inc., under cooperative agreement
with the National Science Foundation.} {\it echelle} package.
The bias and overscan were subtracted. The wavelength calibration was 
derived from Th-Ar data. Flatfields were
taken through a diffusor slide to create ``milky flats'' that made the orders
sufficiently wide to get good flatfields along the edges of the orders
of the data frames.  
Parts of orders with overlapping wavelength coverage were coadded together
before analysis. These overlap regions are larger for the bluer
parts of the spectrum. Over the wavelength region where individual
lines were measured (5300-8000\AA), the signal-to-noise (S/N) is $\sim$ 30. The CN and CH
bandhead regions had lower S/N (S/N$\sim$10).

\section{Abundance Analysis}

We analyzed both the spectrum of \moaspaces and the spectrum of the
Sun \citep{kursol} using the same set of lines, line parameters, and model
atmosphere grids. 
We used TurboSpectrum \citep{turbo}, a 1-dimensional LTE code, 
to derive abundances.
TurboSpectrum
uses the recent treatment of damping from \citet{barklem:00}. 
We interpolated the
model atmosphere grid of ATLAS9 models\footnote{http://kurucz.harvard.edu/grids.html} updated with
new opacity distribution functions \citep{castelli:03}. 
The abundances of most elements 
were determined from analysis of the equivalent widths (EWs). The
EWs for both \moaspaces and the Sun are presented in Table 1.  We restricted the
analysis to lines with EW$\leq 150$m\AA. The EWs were
measured using SPECTRE\footnote{http://verdi.as.utexas.edu/spectre.html} 
(C. Sneden, 2007, private communication). We compared synthetic with observed spectra
to determine abundances for lines that were blended, lines
that had substantial hyperfine splitting and for C and N that
were measured from CH and CN lines, respectively. We used
the solar atlas of \citet{charlotte} to check for blending
with telluric features and eliminated the few lines that  
were affected.

The linelists for CH and CN are from B. Plez (2006, private communication).
The effect of
hyperfine splitting (HFS) was included for Sc, Mn, Cu and Ba. The
HFS constants were taken from the sources listed in \citet{johnson:06}. HFS
information was not available for the Na or Al lines, so neither
this study nor the literature studies we use for comparison can correct for
those effects. Therefore, the comparison between  [Na/Fe] and [Al/Fe] values
for \moaspaces and the literature values is robust, but the absolute
values of these abundance ratios for all studies have a systematic error. 
Table 1 lists the transition probabilities (listed as log $gf$-values) 
and sources for all the atomic lines.  

\subsection{Atmospheric Parameters}

\begin{figure}
\includegraphics[width=16cm]{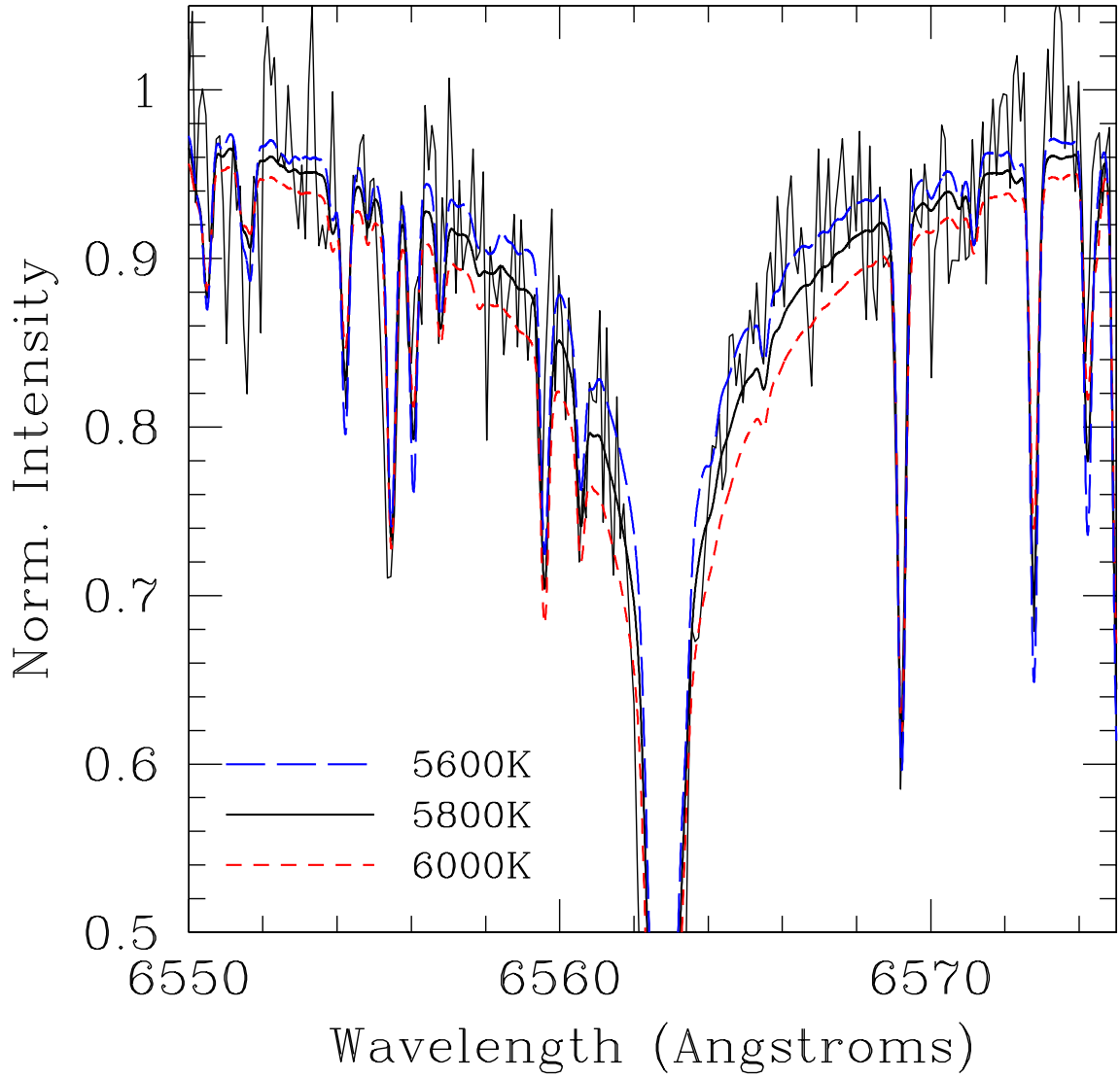}
\figcaption{Fits to the H-$\alpha$ line for three different temperatures:
5600K, 5800K and 6000K. The atmosphere with \teff=5800K is the best
fit to the hydrogen line.}
\label{fig:halpha}
\end{figure}

\begin{figure}
\includegraphics[width=16cm]{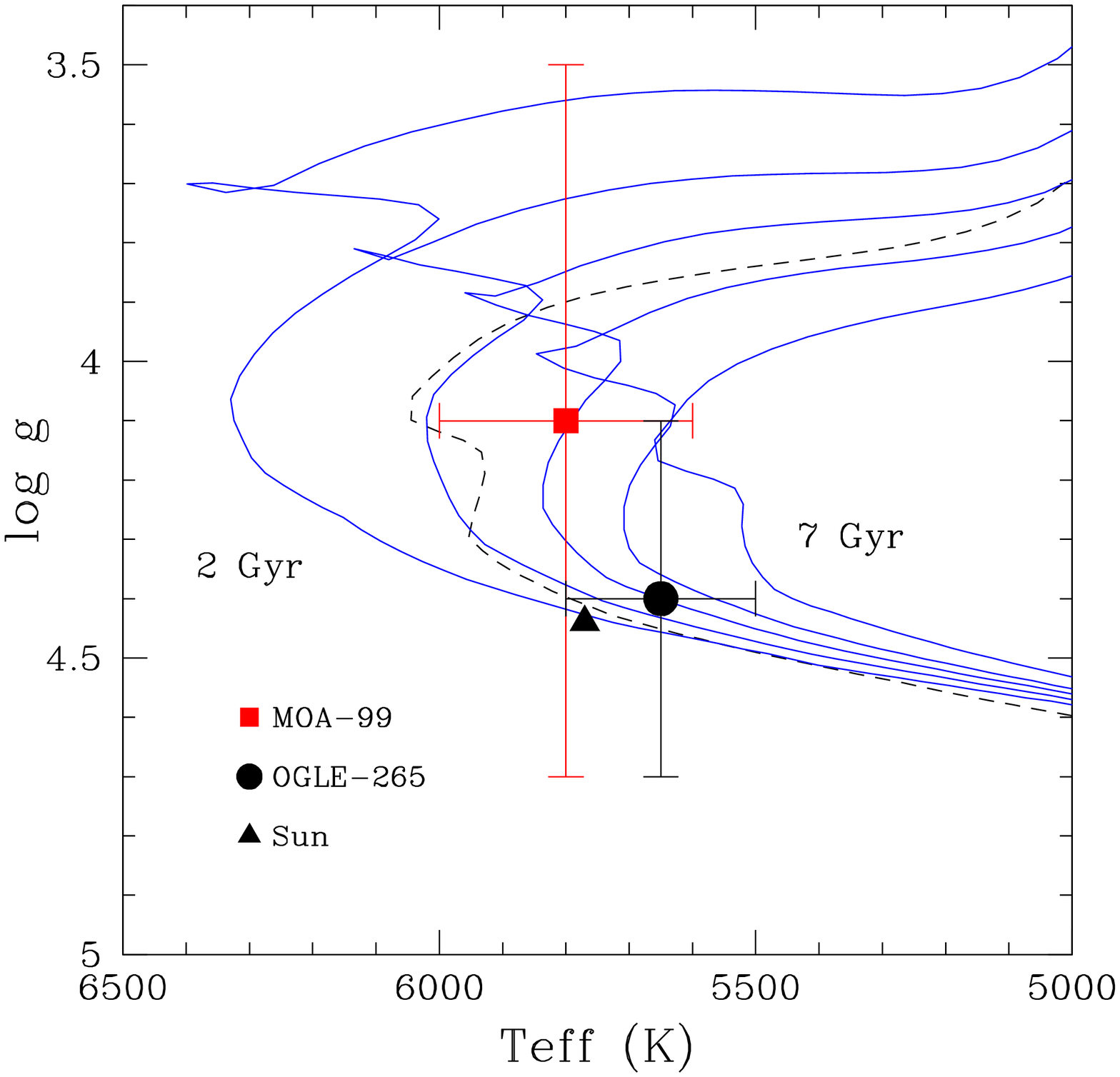}
\caption{The position of \moaspaces (square), \oglespace (circle)
and the Sun (triangle) in the H-R diagram. We also show
isochrones from \citet{yi:01}. The solid lines show isochrones for
[Fe/H]=0.385 and for ages 2, 3, 4, 5 and 7 Gyr. The dashed line
shows a 5 Gyr isochrone for a solar metallicity.}
\label{fig:iso}
\end{figure}

We measured a \teff=5800K using the wings of the H-$\alpha$ and
H-$\beta$ lines (Figure~\ref{fig:halpha}). Our uncertainty in the
temperature is based on the uncertainty in this fit, in particular the
uncertainty in the level of the continuum. Temperatures derived from
the Balmer lines for metal-rich dwarfs show good agreement with
temperatures derived from the infrared flux method and from $V-K$
colors \citep{mashonkina:99,barklem:02}.  For \moas, we compared the
temperature derived from the Balmer lines with that derived from
excitation equilibrium of the \ion{Fe}{1} lines. The temperature
derived from the \ion{Fe}{1} lines was very uncertain because the
abundances derived from individual lines were poorly determined because of
 the S/N
of the spectrum.  The excitation equilibrium favored a higher
temperature (+200K) with an uncertainty of 300K.Given 
the uncertainties, this temperature is in agreement with the Balmer
line temperature. Next, we measured the microturbulent velocity, $\xi$, 
by ensuring that the abundances derived from the \ion{Fe}{1} lines
do not depend on their reduced EW. Our best fit value was $\xi$=1.5 km/s.
Changing $\xi$ by $\pm$0.3 km/s gave marginal fits to the data, 
and we adopt that as our uncertainty
in $\xi$. The gravity was measured by
ionization balance for \ion{Fe}{1} and \ion{Fe}{2} and for \ion{Ti}{1}
and \ion{Ti}{2}. As a sanity check, in Figure~\ref{fig:iso}, we compare the
atmosphere parameters for \moas, \ogle, and the Sun to the parameters from
the Yonsei-Yale \citep{yi:01,demarque:04} isochrones.
Finally, the metallicity of the model atmosphere, [m/H], was set equal to the
[Fe/H] given by the \ion{Fe}{1} lines. Because the \logg{} of the model
atmosphere affects the [m/H] and the [m/H] affects the abundances (and
therefore the \logg{} measurement), we iterated to find a solution for which 
\ion{Fe}{1} and \ion{Fe}{2} were equal and the [m/H] of the model was
equal to [\ion{Fe}{1}/H].

Using standard microlensing techniques (e.g., \citealt{ob03262}),
$\mu$FUN determined that the dereddened color and magnitude of the
source were $(V-I)_0=0.69\pm 0.05$, $I_0=18.17\pm 0.10$. The error
is due to possible differential reddening between the microlensed
source and the red clump, which is assumed to have the same 
$(V-I)_0=+1.00$ as the local Hipparcos clump. If (as seems
likely) the source lies at approximately the Galactocentric distance,
then its absolute magnitude is $M_V=4.5$ and its radius is 1.2$\rsun$. 
That is, it is a solar-type star.
We combined the $\mu$FUN color and the \citet{ramirez:05} 
color-temperature relation (an update of the \citet{alonso:96} relation) to 
derive an estimate of the temperature of 5806$\pm$ 200 K. This photometric
temperature agrees with the temperature from the Balmer lines. For 
\ogle, the spectroscopic and photometric temperatures were different,
suggesting possible systematic errors in estimating the dereddened
color and magnitude of the source star. As outlined in the Introduction,
quantifying any systematic errors is important for obtaining the
most accurate information about microlensing events, and additional data
will be crucial for quantifying any systematic error.

\subsection{Error Analysis}

Our uncertainties are 200K for \teff{} and 0.3 km/s for $\xi$.  The
distribution of EWs and excitation potential for the \ion{Fe}{1} lines
were sufficiently uncorrelated that $\xi$ did not depend on
\teff. Our measurements  of \teff{} and $\xi$ therefore did not depend
on accurate assessments of the other model atmosphere parameters, namely
\logg{} and [m/H]. Our measurement of \logg{} and [m/H], on the other hand, 
depended strongly on the other model atmosphere parameters. 
Because we set the gravity by ionization equilibrium, \logg{}
depends on \teff, $\xi$, and [m/H], as well as the abundance measured
from the \ion{Fe}{1} and \ion{Fe}{2} lines. Therefore, the uncertainty
in \logg{} depends on the uncertainties in those quantities. The
uncertainty in [m/H], in turn, depends on the uncertainty in 
\teff, \logg, and $\xi$, as well as  the scatter in the abundance
given by different \ion{Fe}{1} lines for a particular model
atmosphere. The standard error of the mean for the gravity derived
from the 35 \ion{Fe}{1} lines for a single model atmosphere was 0.04
dex, and for the 4 \ion{Fe}{2} lines was 0.14. We adopt 0.04 dex as
the uncertainty from EW and log $gf$ errors for \ion{Fe}{1} for
inclusion in the [m/H] uncertainty. Adding the \ion{Fe}{1} and
\ion{Fe}{2} uncertainties in quadrature give us 0.15 dex as our
uncertainty in the difference between the \ion{Fe}{1} and the
\ion{Fe}{2} log$\epsilon$\footnote{log~$\epsilon$(A) $\equiv$ {\rm
{log}}$_{\rm 10}$(N$_{\rm A}$/N$_{\rm H}$) + 12.0} values arising from
the EW and oscillator strength uncertainties. The total uncertainty in
\logg{} is 0.66 dex and in [m/H] is 0.21 dex.

We ran the \ion{Fe}{1} and \ion{Fe}{2} EWs through
a series of models: $\pm$ 200K, $\pm$ 0.3 km/s, $\pm$0.3 dex for \logg, and
0.13 dex in [m/H] (smaller because of the limits of the Kurucz
grid), and calculated the difference in Fe abundance with these
different model atmospheres.

Finally, we calculate uncertainties using modified equations (A5) 
and (A20) from
\citet{mcwilliam:95}.
We considered the covariance between \teff : \logg , \teff : [m/H] , 
\logg : [m/H] , $\xi$ : \logg , and $\xi$ : [m/H]. The covariances
were calculated by a Monte Carlo technique. For example, to
calculate the covariance between \teff{} and \logg, we first found
$\partial\log g/\partial{\rm T_{eff}}$ and noted the remaining scatter that was
caused by uncertainty in $\xi$, etc. Next, we randomly picked 1000 \teff{}
from a Gaussian distribution with a $\sigma$ of 200K. We used the derivative
to calculate \logg{} and then added an extra random $\Delta$\logg drawn
from a Gaussian distribution with a $\sigma$ equal to the uncertainty
from non-\teff{} causes. We calculated the covariance
using these 1000 \teff-\logg{} pairs. A similar calculation was done for
the other covariances.

\section{Results}

In Table~2, we summarize the abundances measured for 17 elements
in \moas. We include both log$\epsilon$ and its error, 
as well as [X/Fe] and its error.
To give an idea of the uncertainty due to scatter from the lines, rather
than from atmospheric parameters, we give $\sigma$, the rms of abundances
derived from individual lines as
well as the number of lines. We also give our measurements of the
solar abundances, which we will use to calculate ratios. For reference,
we include the \citet{grevsauv} solar abundances in the final column.

\subsection{Metallicity}

\begin{figure}
\includegraphics[width=16cm]{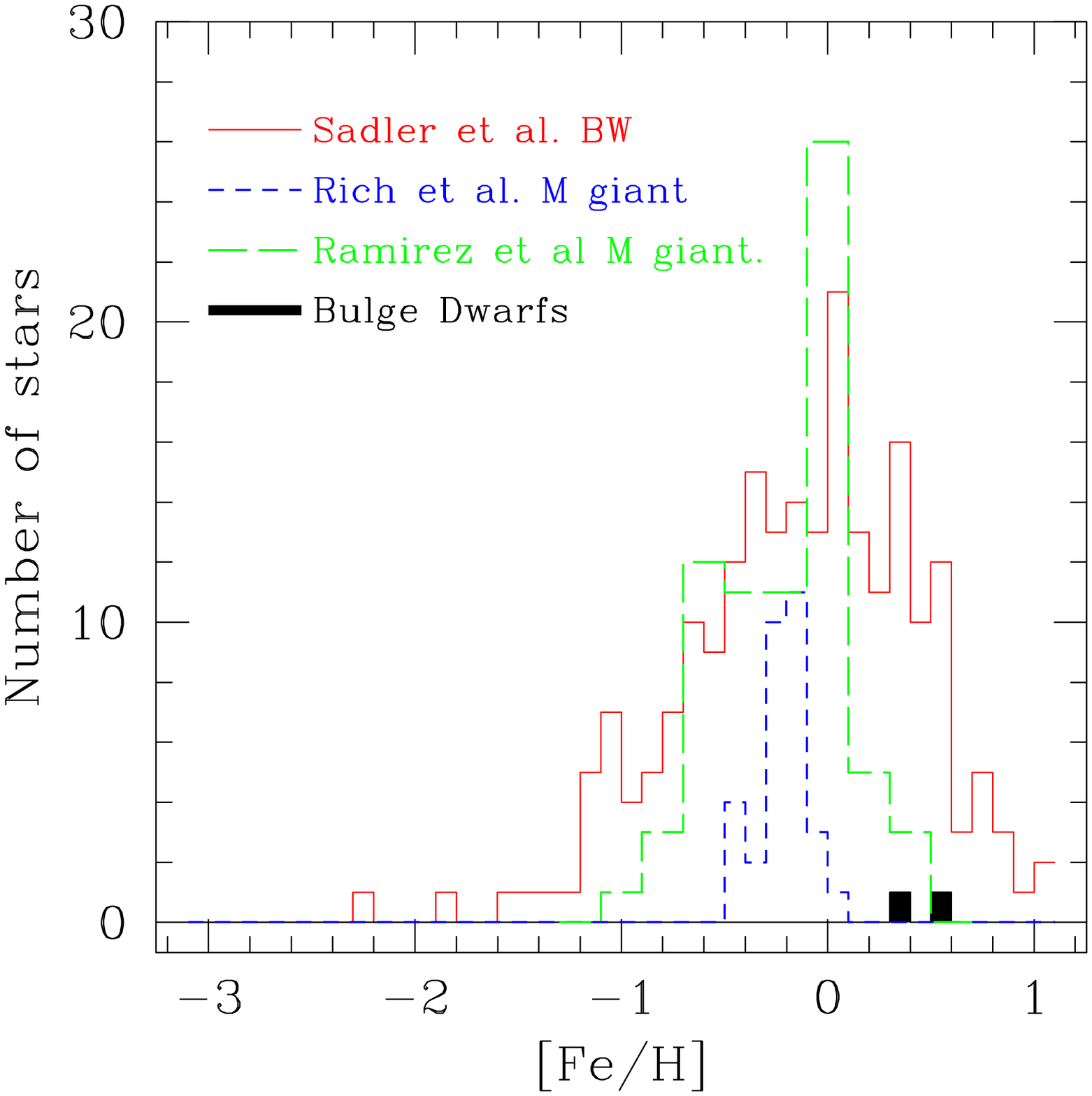}
\caption{The MDF for \moaspaces and \oglespace compared to the 
MDF from \citet{sadler:96}, which was measured on K and M giants and
with the MDFs of \citet{ramirez:00}, \citet{rich:05}, and \citet{rich:07} for
M giants. The MDF of the dwarfs is shifted to higher metallicities compared
to the M giants, which is surprising since the most metal-rich stars
should end their red giant phase as M giants rather than K giants.
}
\label{fig:mdf1}
\end{figure}

\begin{figure}
\includegraphics[width=16cm]{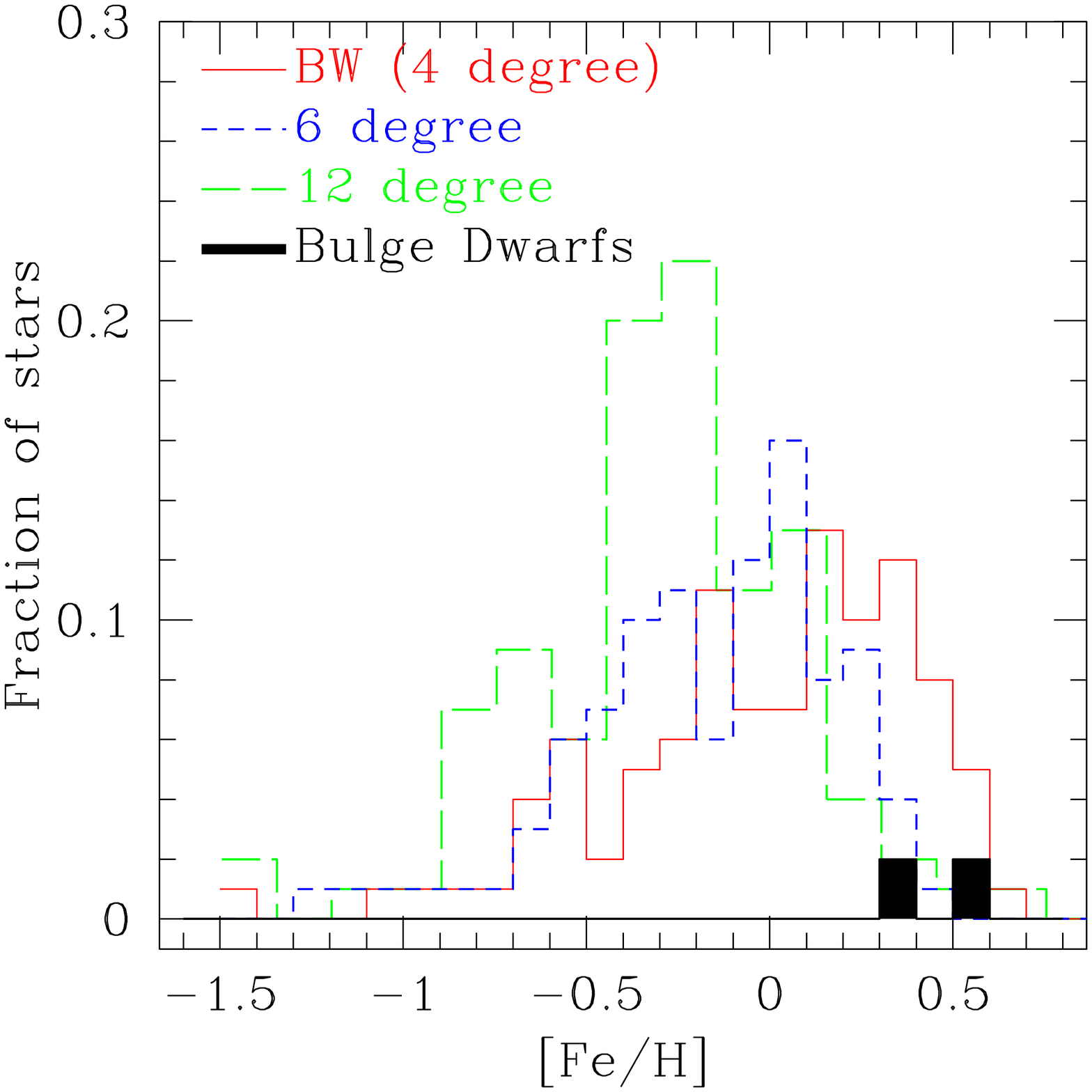}
\caption{The MDF for \moaspaces and \oglespace compared to the MDF
derived by \citet{zoccali:08} from high-dispersion spectra of
giants in three fields: Baade's Window at 4\deg{} as well as 
a 6\deg{} and 12\deg{} field. The fraction of the two-bulge-dwarf sample
has been scaled down from 0.5 to fit clearly on the graph. The average metallicity
of the bulge giants decreases as the distance from the Galactic center. The
dwarfs are more than 4\deg{} away from the center, making their high metallicities
even more surprising. }
\label{fig:mdf2}
\end{figure}

We measure [Fe/H]$=0.36\pm0.18$ for \moas. In \citet{johnson:07}, 
we measured [Fe/H]$=0.56\pm0.19$ for the dwarf \ogle.
The stars that are microlensed are unbiased in metallicity. The criterion
for spectroscopic follow-up is that the unmagnified source be faint enough to
be a Bulge dwarf, regardless of color. Therefore, especially 
considering the large and variable reddening toward the Bulge, we are
not biased in our high-resolution follow-up toward high metallicity
sources. The high metallicities of these two dwarfs is surprising
given that work on giants has indicated an average metallicity
near solar. In Figure ~\ref{fig:mdf1}, we compare the metallicity distribution function
(MDF) of the two dwarfs with several MDFs based on studies of giant
stars. The MDF of \citet{rich:07}, which is based on high-resolution analysis of M giants and should be
biased {\it toward} the highest metallicity objects, lacks any giants as
metal-rich as these dwarfs. This is particularly notable, since the
work of \citet{sadler:96} on K giants with low-dispersion spectra shows some 
extremely metal-rich stars. A complicating factor is the possible presence
of a metallicity gradient in the Bulge. In Figure~\ref{fig:mdf2}, we compare the
metallicities of the dwarfs with MDFs derived by \citet{zoccali:08} 
from high-resolution spectra for giants in three Bulge fields: 4\deg, 6\deg,
and 12\deg away from the Galactic center. The inner field is more metal-rich
than the outer. However, these two dwarfs are located 6.5\deg (\moas) and 
4.9\deg (\ogle)
away from the Galactic Center, and therefore gradients cannot explain their
anomalously high metallicities. 

These results hint that the MDFs of the Bulge giants and dwarfs may be different. 
Whether this is true can be established by more observations of Bulge dwarfs and by
resolution of the discrepancies among the MDFs derived for giants, particularily
between the low-dispersion and high-dispersion studies. Because the microlensed dwarfs
are found at a range of distances from the Galactic Center, comparison of the giant
and dwarf MDFs also depends on measuring the metallicity gradient (and the size of deviations
from that gradient) in the Bulge. Ideally, the MDF for giants in the same field as the microlensed
dwarf would be measured.

\subsection{Comparison with Isochrones and the Age of \moas}

\begin{figure}
\includegraphics[width=16cm]{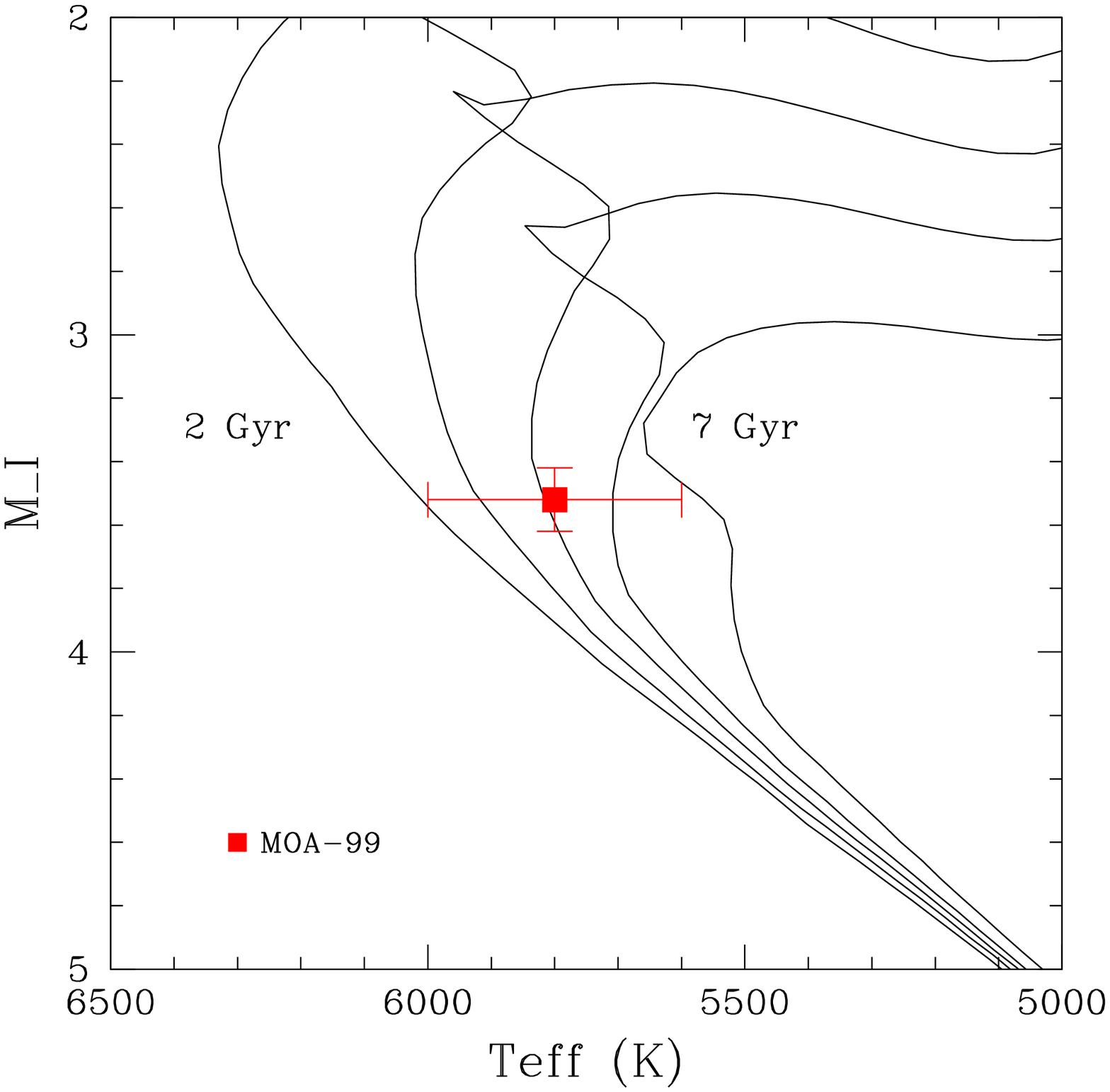}
\caption{The position of \moaspaces (square) in the H-R diagram
(M$_I$-\teff). M$_I$ was calculated using the I$_0$ magnitude and
assuming a distance of 8.5 kpc. The \teff{} is the temperature from
the Balmer lines. We also show
isochrones from \citet{yi:01} The solid lines show isochrones for
[Fe/H]=0.385 and for ages 2, 3, 4, 5 and 7 Gyr. The dashed line
shows a 5 Gyr isochrone for a solar metallicity.}
\label{fig:imag}

\end{figure}

With our measurements of \teff, \logg, and [Fe/H] from spectroscopy, we 
can compare the position of \moaspaces on the Hertzsprung-Russell diagram with 
theoretical isochrones (Fig.~\ref{fig:iso}). 
We use the Yonsei-Yale isochrones
\citep{yi:01,demarque:04}
with [Fe/H]=0.385, [$\alpha$/Fe]=0 for comparison. The best fit
age for this star is $\sim$5 Gyr. Instead of log g, we can
also use the I-band magnitude to plot the star on the H-R diagram. 
We adopt a distance of 8.5 kpc, placing this
dwarf on the far side of the Bulge, its most likely position because the
optical depth for lensing is larger there. Figure~\ref{fig:imag} shows that a similar age
is obtained. Indeed,
shifting the position of \moaspaces in the vertical direction, by changing its
distance or luminosity has little effect on the young age 
that we derive for this star, 
because, at this metallicity, there should be no stars this hot with ages $\geq$ 6 Gyr.
However, the uncertainty in the temperature combined with the effect
that changing the temperature has on the derived metallicity produce large
uncertainties in the age. If we instead adopt the T$_{\rm eff}$
on the lower edge of our range (5600K), the metallicity calculated
from the \ion{Fe}{1} lines drops to [Fe/H]=0.16 dex. Using isochrones
of this metallicity, the new \teff{} gives an age of $\sim$9 Gyrs. Improvements
in the accuracy of temperatures are needed to get better age constraints for
Bulge dwarfs, but in principal, ages can be measured for individual stars near the
main-sequence turnoff.

\subsubsection{Mixing in Giants in the Bulge}

\begin{figure}
\includegraphics[width=16cm]{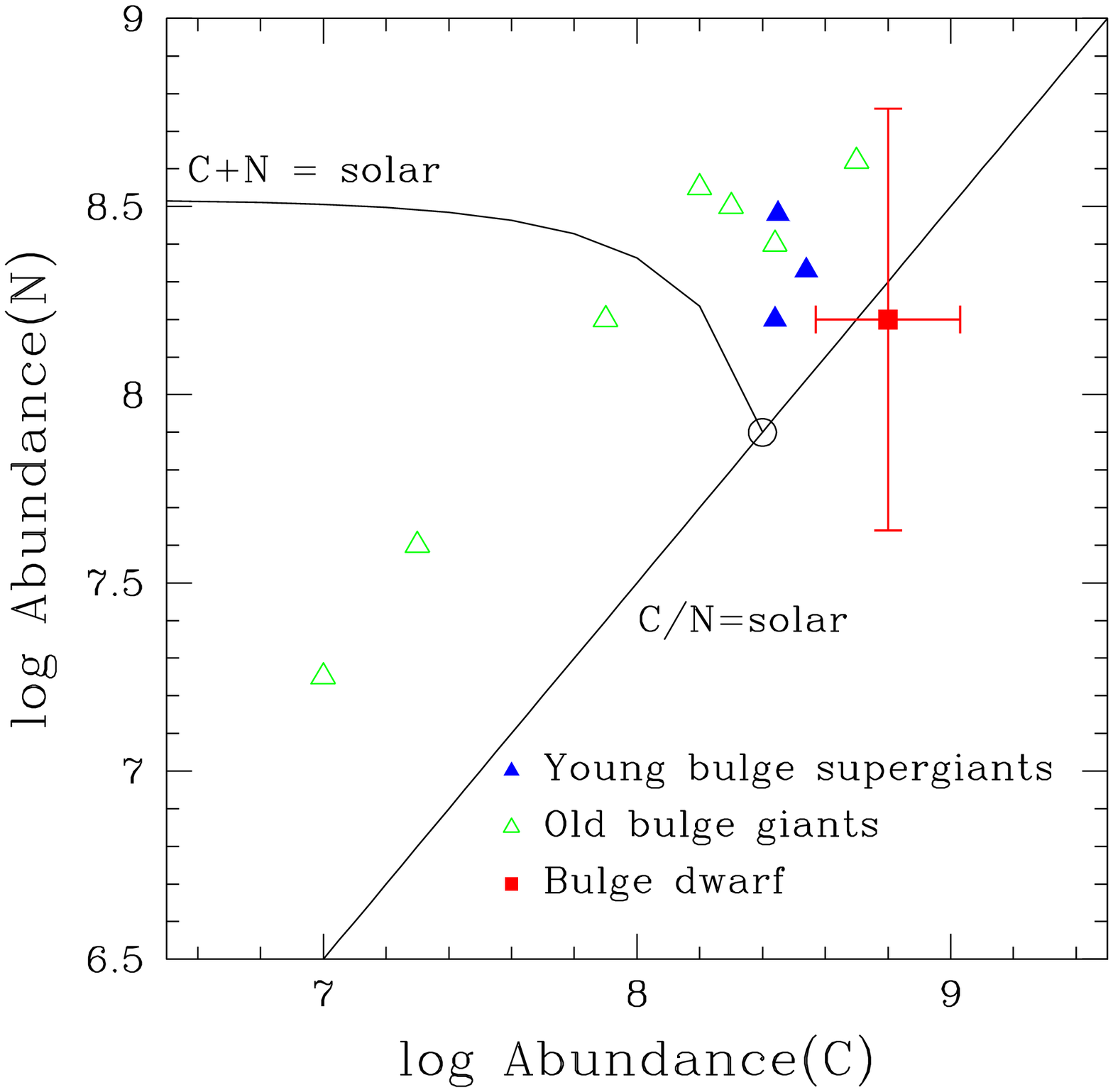}
\caption{The C and N abundance of \moaspaces (filled red square) 
compared with giants in the Bulge. Open green triangles show old giants 
from \citet{cunha:06} and filled blue triangles show young supergiants
from \citet{cunha:07}. The open black circle shows the position of the
Sun and the straight solid line marks where C/N ratio is solar. The 
curved line represents constant C+N. The C and N abundances in
the giants can be explained by conversion of some C to N in CN processing
from the solid line. The data for \moaspaces show that the solid
line is a reasonable representation of the ``primordial'' (unaffected by
internal mixing) C/N ratio in the Bulge. 
}
\label{fig:mix}
\end{figure}

As stars move up the giant branch, they pass through first dredge-up, 
which brings up material that has been processed in the CN cycle. 
The C and N
abundances measured in giants no longer represent the original C and
N endowments of the stars, although C+N will remain constant as long
as only material processed in the CN cycle, and not the ON cycle, is mixed
to the surface. 
\citet{cunha:06} and \citet{cunha:07} measured C and N in giants
in the Bulge. They found that
the giants lie to the N-rich side of the line defined by the C/N ratio
of the Sun (Figure~\ref{fig:mix}).
\citet{cunha:07} concluded that a small
amount of mixing had occurred in the giants. This conclusion is only valid if
the original abundances in the giants lie close to the line. Otherwise, 
if the Bulge dwarfs have non-solar
C/N ratios, the C/N ratios measured in Bulge giants 
could imply either no mixing (and a N-rich original composition) 
or substantial mixing (and a C-rich original composition). 
The abundances of C and N for \moaspaces are also plotted on 
Figure~\ref{fig:cn} and shows that the assumption of \citet{cunha:07} 
is justified
and that the expected amount of mixing has occurred in the giants.

\subsection{Lithium}

\begin{figure}
\includegraphics[width=16cm]{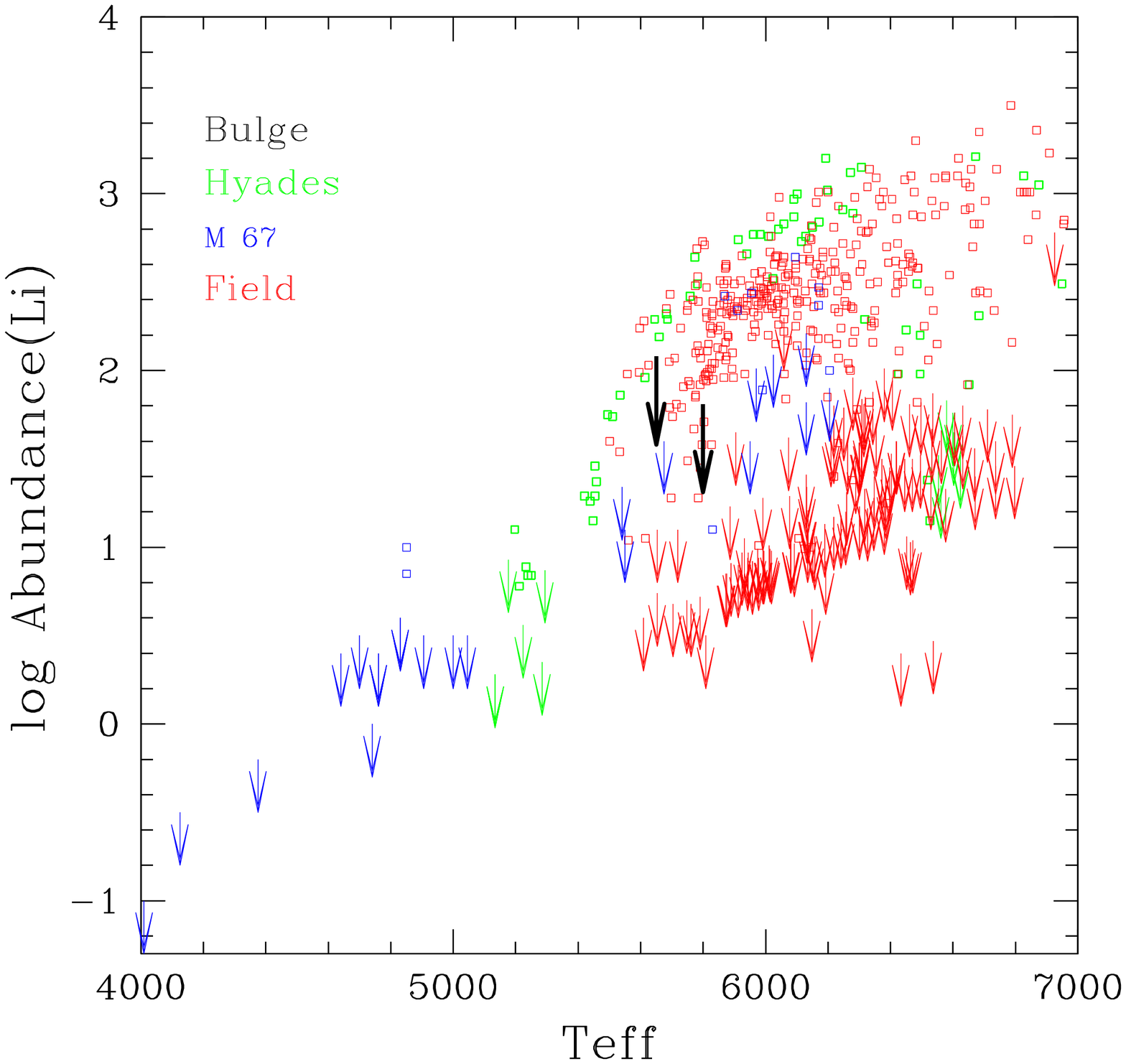}
\caption{\teff vs. log$\epsilon$(Li) for the Hyades and M67 from
\citet{balachandran:95} and for field stars from \citet{lambert:04}
compared with the upper limit for \moaspaces and \ogle. The bulge dwarf
Li limits are consistent with disk stars, which is not surprising given the
age and temperature of these stars.}
\label{fig:li}
\end{figure}

Li has been created since the Big Bang by stellar nucleosynthesis and by cosmic
ray spallation. A Li abundance for a star in the Bulge, measuring
how fast Li was made in the early Galaxy, would be very interesting. However,
most stars no longer have the same amount of Li on their surfaces as was present in their
natal gas clouds. Li is easily burned during pre-main sequence and main-sequence phases
of stars and is either destroyed throughout the convective envelope during the RGB phase or
(for a brief time) created in the star itself and dredged up.
We have no detection of Li in this star, only a 3-$\sigma$ upper limit
of log$\epsilon$(Li)=1.84 dex based on a $\chi^2$ fit to the data
(see \citealt{johnson:07} for more details).  In Figure~\ref{fig:li} we show this
upper limit compared with Li measurements in open cluster stars having 
a range of ages as well as field stars from \citet{lambert:04}. We
also include the upper limit from \ogle. Lower values can be 
expected in field stars because of astration on the main-sequence and because
they are often older than the clusters featured in Figure~\ref{fig:li}
and were formed out of
gas that had not been polluted by as much Li. Figure~\ref{fig:li} shows that
the Li upper limits in the bulge dwarfs are consistent with the upper limits in
field dwarfs. A dwarf with \teff$>7000$K is probably needed to measure the amount
of Li produced by spallation in the Bulge.

\subsection{Chemical Evolution of the Bulge}

The Bulge has a different star formation history than the halo/disk. The 
ratios of Type II/Type Ia pollution or Type II/AGB star pollution at
a given [Fe/H] are therefore different as well, and the abundance
ratios reflect this.
We compare the abundances for both \moaspaces and
\oglespace with Bulge giants and field stars from the thick/thin
disk and halo from literature sources. In Table 3, we summarize the
literature sources we use for each element.

\subsubsection{Carbon and nitrogen}

For many elements, observing red giant stars in the Bulge is an
effective method of measuring their abundances. However, as shown in
\S 5.2.1, internal
mixing on the RGB alters the abundances of
C and N. The abundances of C and N that we measure
in \moaspaces therefore represent the first observations of the primordial
C and N produced by the chemical evolution of the Bulge.

\moaspaces has [C/Fe]=0.04$\pm0.22$ and [N/Fe]=$-$0.06$\pm0.43$ (Fig.~\ref{fig:cn}).
The solar values of [C/Fe] and {N/Fe] show that C and N production
kept pace with the Fe production in the Bulge. There are many sources
of C in the Universe \citep[e.g.]{gustafsson:99}.  Type II SNe and
AGB stars certainly contribute substantial amounts of C and N; the
roles of novae and Wolf-Rayet stars are less clear. These contributions,
whatever they are, track the production of Fe in the
chemical evolution of the Bulge.  Finally, we
attempted to measure the $^{12}$C/$^{13}$C ratio, which is
sensitive to the source of C, being low for low-mass AGB stars and
high for Type II SNe. We could only set
an uninteresting limit ($^{12}$C/$^{13}$C > 1) on this very interesting
number.

\begin{figure}
\includegraphics[width=16cm]{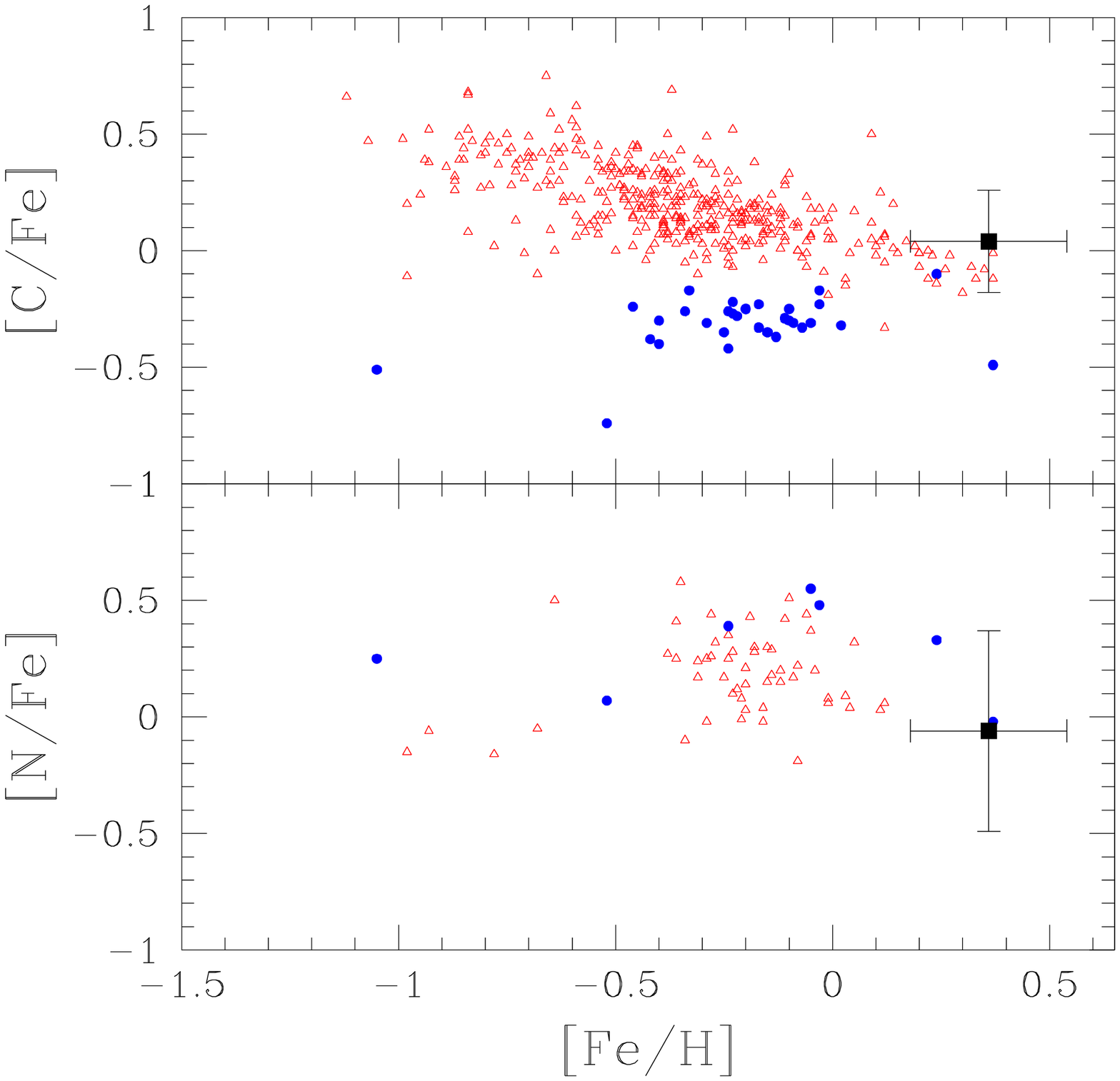}
\caption{[C/Fe] and [N/Fe] for \moaspaces (filled black square) compared to
Bulge giants (filled blue circles)
and disk dwarfs (open red triangles). The low [C/Fe] values for the Bulge
giants are the result of internal mixing. The [C/Fe] and [N/Fe] values in
\moas, on the other hand, are the result of the pollution of the gas of the Bulge by
previous generations of stars. The solar ratios for these elements impose constraints on the
inefficiency of C and N production in the Bulge.}

\label{fig:cn}
\end{figure}

\subsection{Sodium and Aluminum}

The abundances of Na and Al are elevated in Bulge giants
\citep{mcwilliam:94, lecureur:07} compared to disk stars.  Metal-rich
Type II SNe are predicted to produce more of the odd-Z elements
such as Na and Al than metal-poor Type II SNe and could potentially be the explanation of the
difference between the Bulge and the disk. In Figure~\ref{fig:naal},
we show the [Na/Fe] and [Al/Fe] value for \moaspaces and \ogle. These
two unmixed stars are on the lower end of the scatter seen in the
giants. This could be a hint that the larger [Na/Fe] and [Al/Fe]
values seen in giants are due to internal mixing, but because the
dwarf values fall within the scatter outlined in the scatter indicates
that more measurements in dwarf stars are needed before any
differences in the distribution of [Na/Fe] and [Al/Fe] in dwarfs and
giants can be seen.

\begin{figure}
\includegraphics[width=16cm]{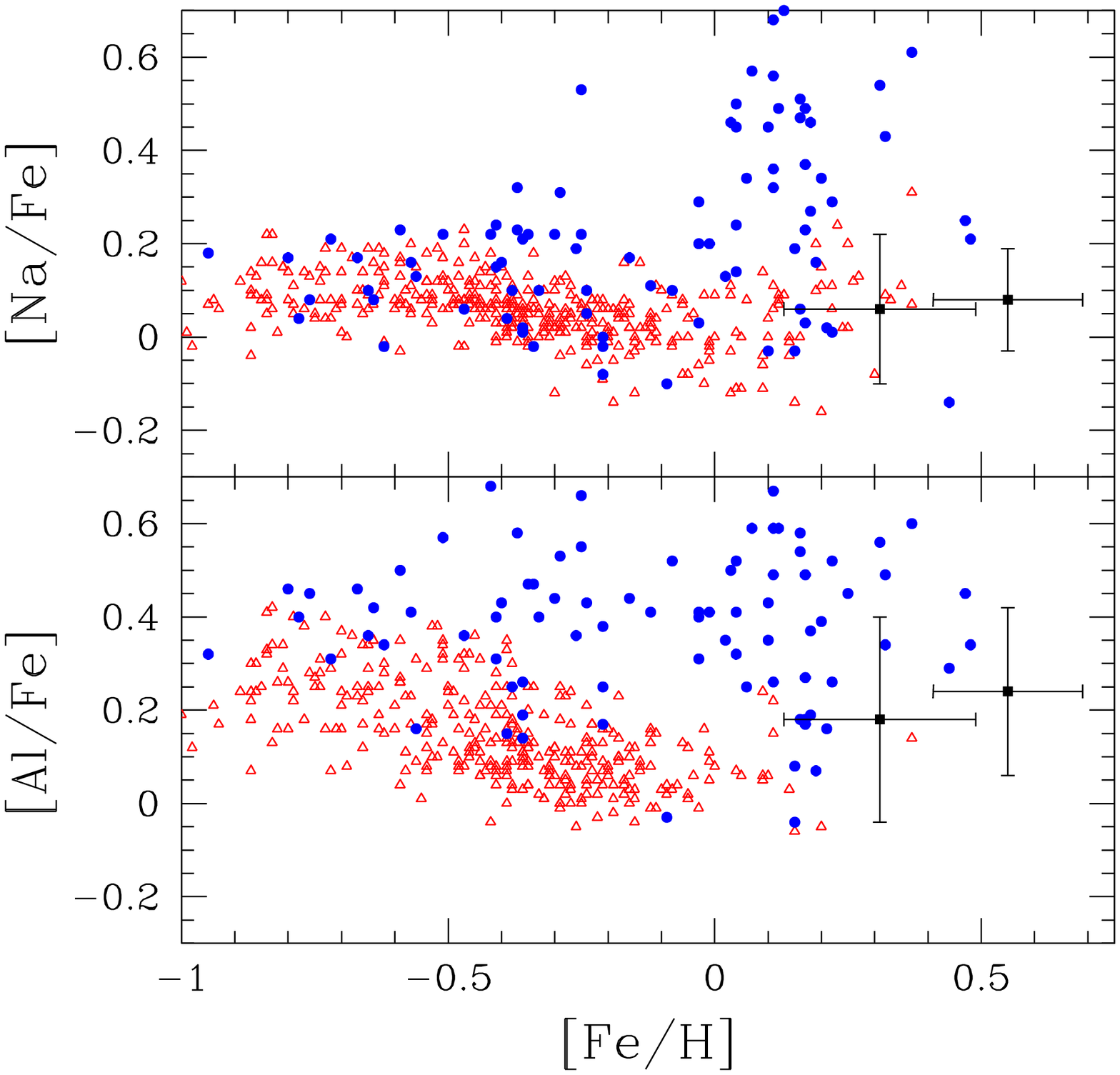}
\caption{[Na/Fe] and [Al/Fe] for \moaspaces and \oglespace (filled black 
squares) 
compared to
Bulge giants (filled blue circles)
and field stars (open red triangles). The dwarfs fall within the distribution of
[Na/Fe] and [Al/Fe] values seen in the giants, but at the lower edge of that
distribution. 
}
\label{fig:naal}
\end{figure}

\subsection{The $\alpha$ elements}
Figure~\ref{fig:alpha} shows the [O/Fe], 
[Mg/Fe], [Si/Fe], and [Ca/Fe] for
\moaspaces compared with halo stars, thin and thick disk stars, and Bulge giants.
The metallicity of \moaspaces is in the range of metallicities measured for
Bulge giants, allowing direct comparison of [$\alpha$/Fe] ratios between
dwarfs and giants. The agreement is good, as both the giants and the
dwarf have [$\alpha$/Fe] below the high [$\alpha$/Fe] values of the 
more metal-poor ([Fe/H]$\leq$0) stars. This decline in all [$\alpha$/Fe]
ratios suggests that Fe from Type Ia SNe is being added and that the more
metal-rich stars formed sufficiently later to have this ejectum in their gas.

\begin{figure}
\includegraphics[width=12cm,angle=270]{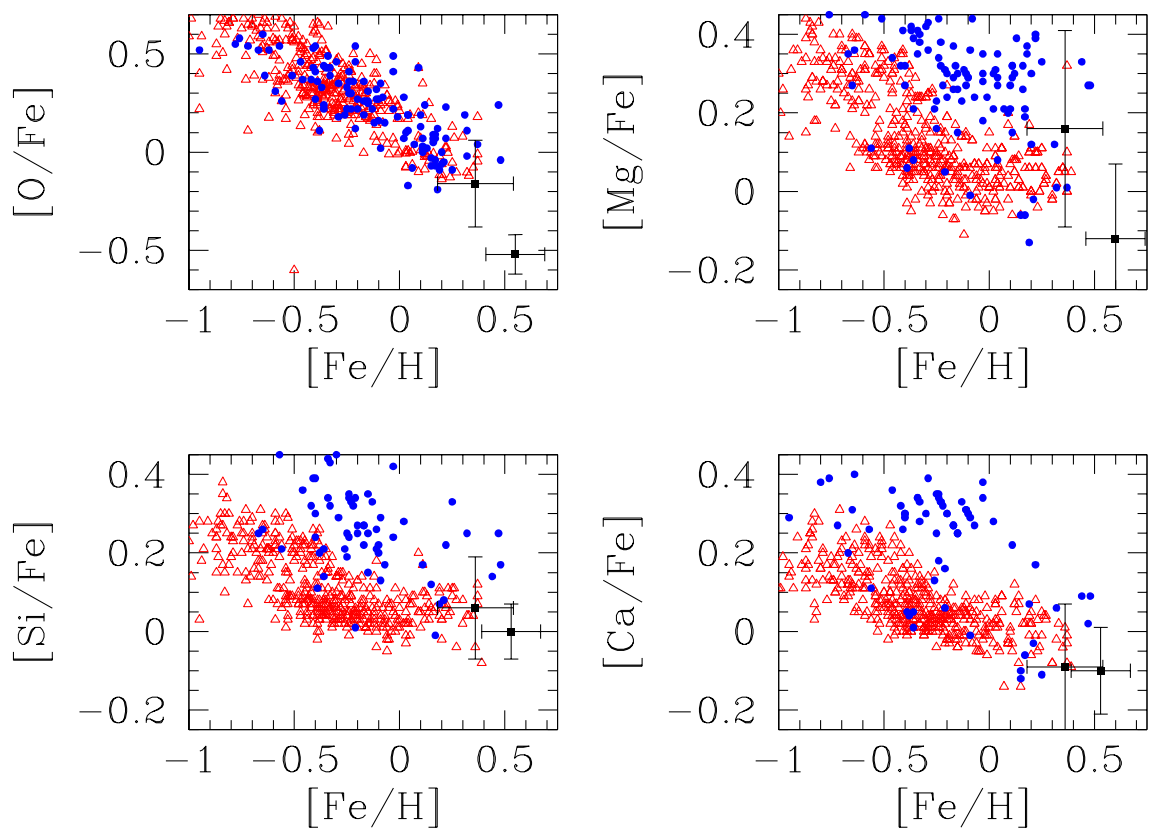}
\caption{[O/Fe], [Mg/Fe], [Si/Fe] and [Ca/Fe] 
for \moaspaces and \oglespace (filled black squares) compared to
Bulge giants (filled blue circles)
and field stars (open red triangles). The [$\alpha$/Fe] ratios
in \moaspaces agree well with the values measured in giants of similar
metallicity. 
}
\label{fig:alpha}
\end{figure}

\subsection{Potassium} K is an odd-Z element and is predicted in
nucleosynthesis models to be underproduced relative to the $\alpha$
elements in metal-poor SNe. There are few measurements 
in the literature, and those that exist show the opposite
trend of increasing [K/Fe] with decreasing [Fe/H] \citep{gratton:87,chen:00,
cayrel:04}. 
However,
the only K line available for study in most stars, the resonance
line at 7698\AA, is affected by non-LTE effects, and these corrections
have not yet been applied to large samples. \citet{zhang:06} derived 
NLTE corrections for each star in their sample. The [K/Fe] values were
still supersolar at low metallicities, with the thin disk stars showing
a drop in [K/Fe] for[Fe/H]$\geq-1$. The [K/Fe] ratios in the
thick disk stars remain high. However, the [K/Mg] ratios showed much smaller
variations among the different Galactic populations. 
They argued that the constant [K/Mg]
ratio ([K/Mg]=$-0.08\pm0.01$) 
in the stars indicated that the nucleosynthesis of
K is closely coupled to that of the $\alpha$-elements, which is somewhat
surprising given the theoretical predictions. 

We measured the 7698\AA{} line
in  \moaspace and the Sun. We also measured the K abundance
in \oglespace using TurboSpectrum and the model atmosphere described
in \citet{johnson:07}. The [K/Fe] we measure for \oglespace is $-0.08\pm0.20$ 
Figure~\ref{fig:k} compares the results for the Bulge to the \citet{zhang:06}
results. 
We applied no NLTE correction, but assumed that
the LTE abundances in the Sun and \moaspaces would be affected by the same
amount. 
 
The [K/Mg] values are 0.07$\pm0.23$ for \moaspaces and 0.12$\pm0.27$ for \ogle; the
[K/Mg] abundance is still within a narrow range, even in this very
different chemical evolution history.

\begin{figure}
\includegraphics[width=16cm]{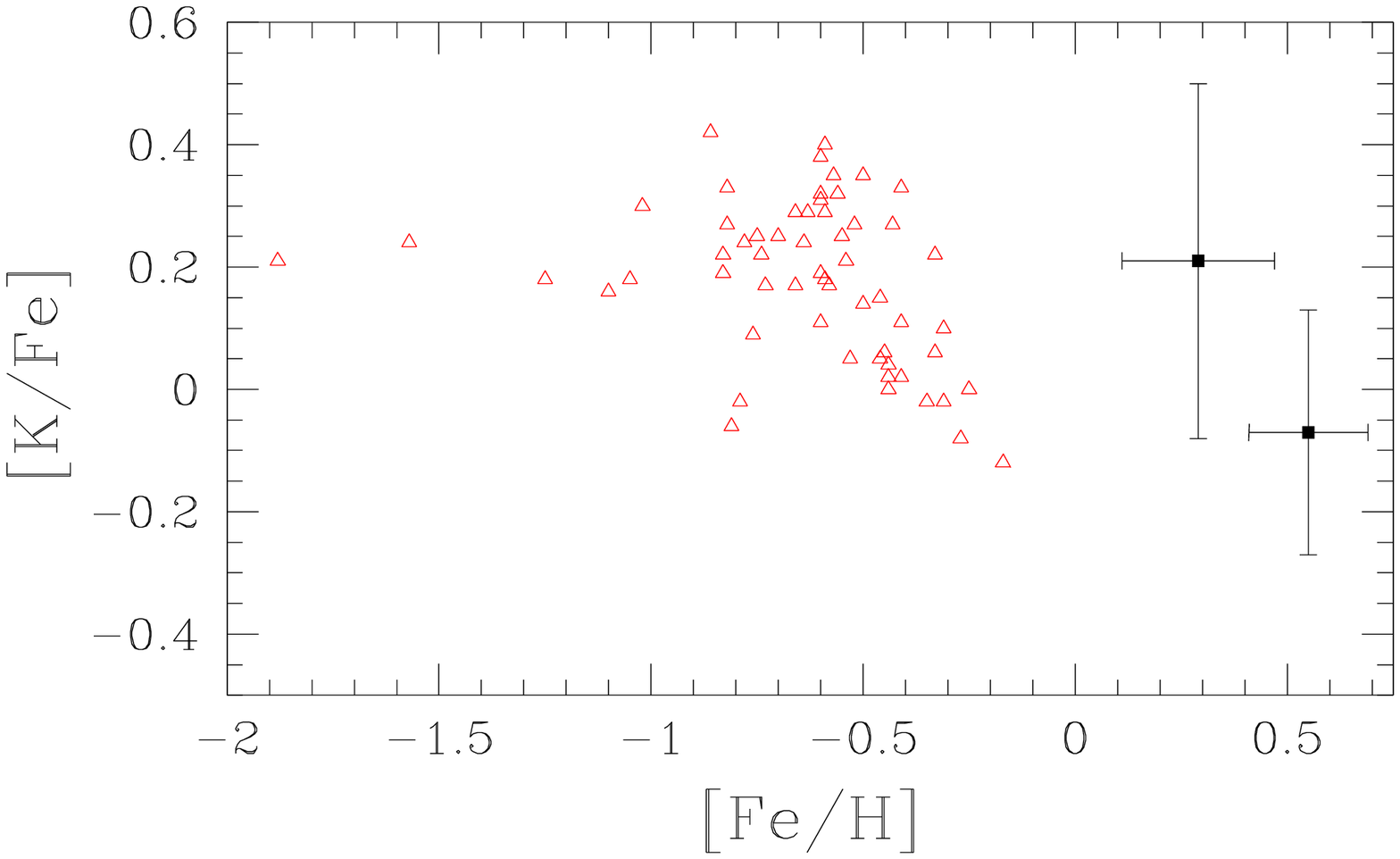}
\caption{[K/Fe] 
for \moaspaces and \oglespace (filled black squares) compared to
field stars (open red triangles). The [K/Fe] values in the
dwarfs fall in the range seen in the disk stars.}
\label{fig:k}
\end{figure}

\subsection{Iron-peak elements}

The abundances for Ti, Sc, Mn and Ni are shown in Figure~\ref{fig:iron}. 
\citet{mcwilliam:94} found that [Ti/Fe] behaves like [O/Fe] in the
Bulge, with supersolar [Ti/Fe] ratios for many stars, followed by
[Ti/Fe] decreasing to solar for [Fe/H}$>0$.  The abundance
of Ti is also enhanced in halo stars, leading it to be classified as
an ``$\alpha$-element'' for observational purposes. In \moas, [Ti/Fe] is 
close to solar, in line with the $\alpha$-elements discussed above. 
The ratios of [Sc/Fe] and [Ni/Fe] are observed to be 
close to solar for a wide range of 
populations: halo, thick and thin disk and Bulge. The data for \moaspaces show
that abundances in this Bulge dwarf agree with this picture. 

For [Fe/H] $<-1.5$ in the Galactic halo/disk, there is a plateau
at [Mn/Fe]$\sim -0.5$ \citep{mcwilliam:95,mcwilliam:03}.  Because Type Ia
SNe have not polluted the most metal-poor stars in the Galaxy, we can
derive the ratio of Mn/Fe produced in (metal-poor) Type
II SNe from this plateau. 
[Mn/O] starts to rise before [O/Fe] starts to drop in the  
disk. Because the drop in [O/Fe] signals the onset of substantial Type Ia SN
contribution, the rise in Mn relative to O cannot be due to Type Ia SNe, but
rather to increased production of Mn by more metal-rich Type II SNe \citep{feltzing:07}.
\citet{mcwilliam:03} also measured Mn in 13 stars in the
Sagittarius dwarf galaxy. These stars have substantial Type Ia SNe contributions
to their gas, but [Mn/Fe] values about 0.2 dex below the trend seen in the
Galactic disk, providing additional evidence that
metal-rich Type II SNe are responsible for Mn production. The near solar [Mn/Fe]
values for \moaspaces and OGLE-99 are also in support of this idea, because they
were seen 
in an environment that 
had many Type II SNe occur, unlike the Sagittarius stars.

\begin{figure}
\includegraphics[width=12cm,angle=270]{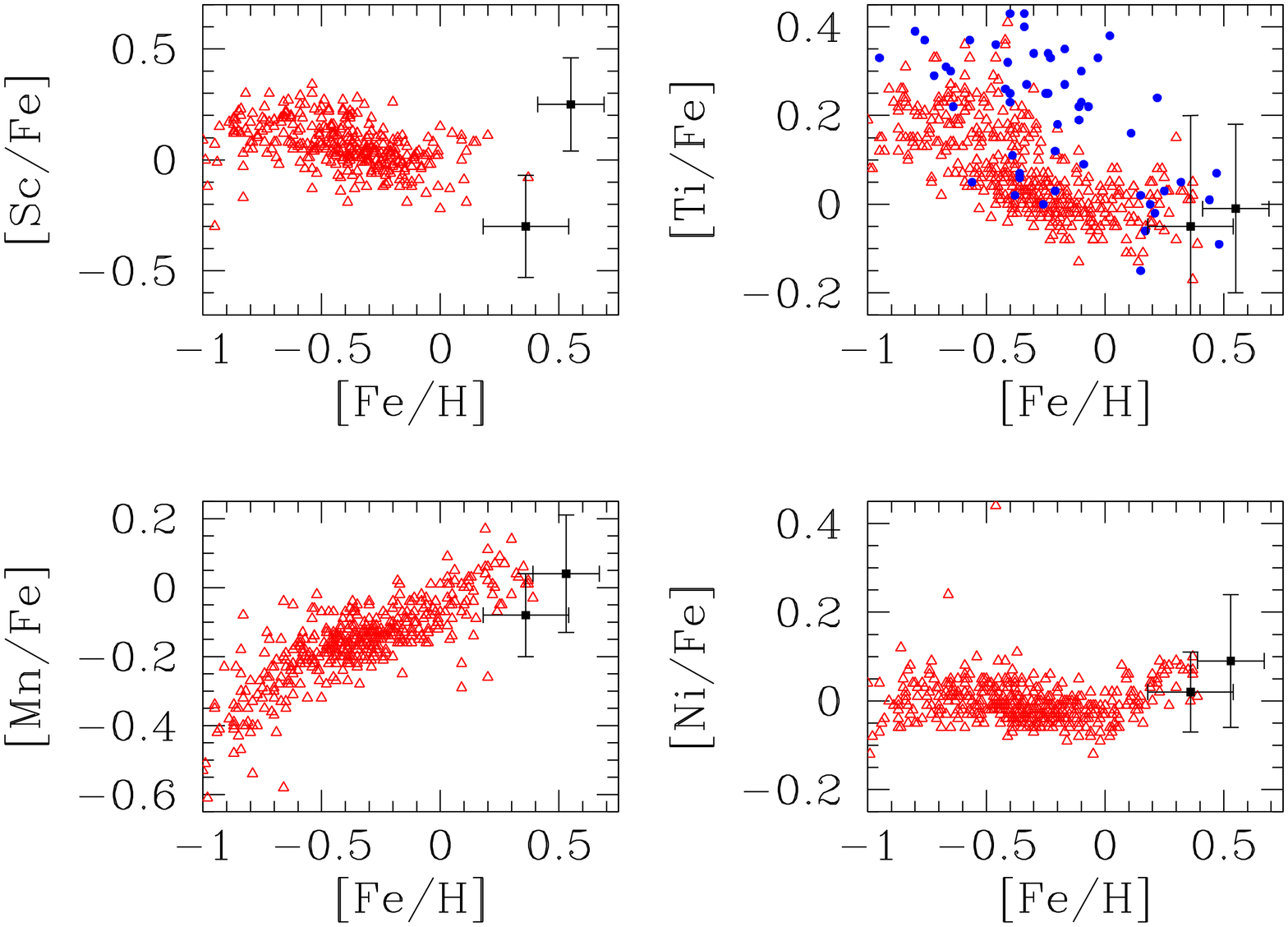}
\caption{[Sc/Fe], [Ti/Fe], [Mn/Fe], and [Ni/Fe] 
for \moaspaces and \oglespace (filled black squares) compared to
Bulge giants (filled blue circles) and field stars (open red triangles).}
The [Ti/Fe] value in \moaspaces agrees well with [Ti/Fe] ratios
measured in bulge giants of similar metallicity. Within the 
error bars, the other [iron-peak/Fe] ratios follow the trends seen
the disk stars.
\label{fig:iron}
\end{figure}

\subsection{Copper and Zinc}

\begin{figure}
\includegraphics[width=16cm]{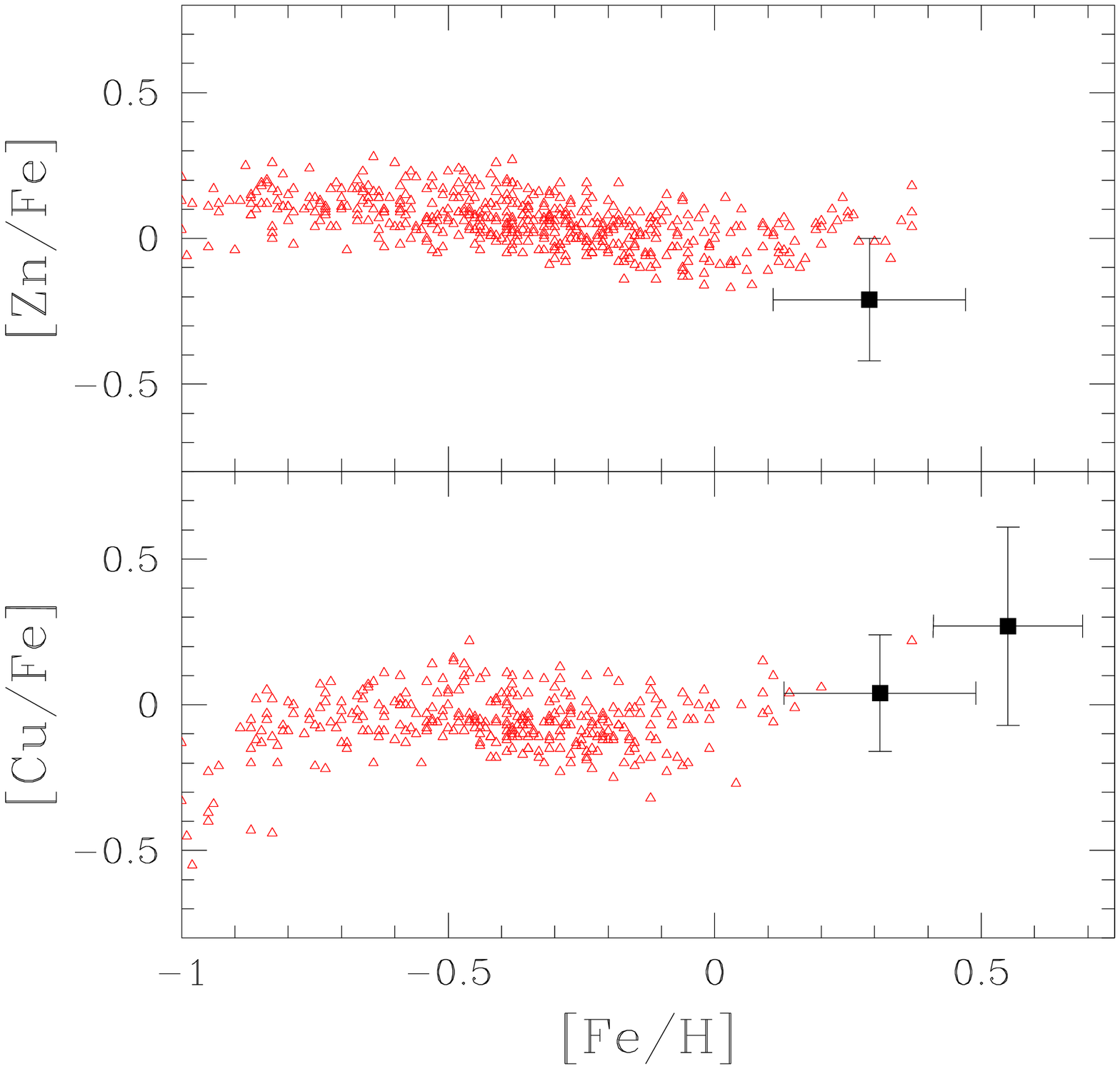}
\caption{[Cu/Fe] and [Zn/Fe] for \moaspaces and \oglespace (filled
black squares) compared to field stars (open red triangles). The \citet{matteucci:99} models predict [Zn/Fe]$\approx0.2$ for high-metallicity stars in
the Bulge. Therefore, the low [Zn/Fe]  
disagrees with the theory that Zn is produced
in large amounts of Type Ia SNe. The solar and supersolar [Cu/Fe] values
in the Bulge are consistent with either Type Ia SN production or
metal-rich Type II SN production.}
\label{fig:cuzn}
\end{figure}

The percent of Cu and Zn production to be ascribed to different
nucleosynthesis sites (e.g., Type II SNe, AGB stars, Type Ia SNe) is
uncertain. The observations show that [Cu/Fe] $\sim -1$ at the lowest
metallicities and then rises to solar by [Fe/H]$\sim -0.8$
\citep[e.g.][]{mishenina:02}. [Zn/Fe], on the other hand, has
supersolar values at the lowest metallicities and
then decreases to closer to solar \citep[e.g.][]{mishenina:02}. 
\citet{matteucci:93} used new weak
s-process calculations and available SN models to argue that
approximately two-thirds of the Zn and Cu production in the Universe
is due to Type Ia SNe. The rest of the Zn is from a primary process in
massive stars, while the Cu comes from a secondary
(=metallicity-dependent) process in massive stars. Using this model
and a chemical evolution model for the Bulge, \citet{matteucci:99}
predicted that both [Cu/Fe] and [Zn/Fe] would be $\sim$ 0.2 at
[Fe/H]=0.3. The SN models available to \citet{matteucci:93} did not
include important effects, such as detailed calculation of
neutron-capture elements beyond Fe. \citet{bisterzo:05} used updated 
results and considered Zn and Cu production by neutron-capture in the
O-rich parts of Type II SNe (``weak sr-process''). In their analysis, Cu
is mostly produced in this weak sr-process, a secondary process. A
small amount of primary Cu is made as radioactive Zn in the inner
regions of Type II SNe.  Zn production is also due to massive star
nucleosynthesis, but here there is a large primary production in the
$\alpha$-rich freeze out in Type II SNe, which is supplemented at higher
metallicities by a secondary contribution from the weak-sr process.
\citet{bisterzo:05} find no need for contributions to Cu and Zn from
Type Ia SNe or AGB stars.

The measurements of [Cu/Fe] and [Zn/Fe] in \moaspaces support the
\citet{bisterzo:05} model and argue against the production of
large amounts of Zn in Type Ia SNe. Because the abundance of
Cu is dominated by a secondary process, the solar [Cu/Fe] at 
[Fe/H]=0.36 and the supersolar value at [Fe/H]=0.56
are the result of copious Cu production in metal-rich Type II SNe. However,
the primary production of Zn and the smaller secondary contribution is
not sufficient to keep up with the Fe from both Type II SNe and Type Ia SNe.
The observations
also show again that the chemical evolution of Zn is separate from Fe.

\subsection{Barium}

\begin{figure}
\includegraphics[width=16cm]{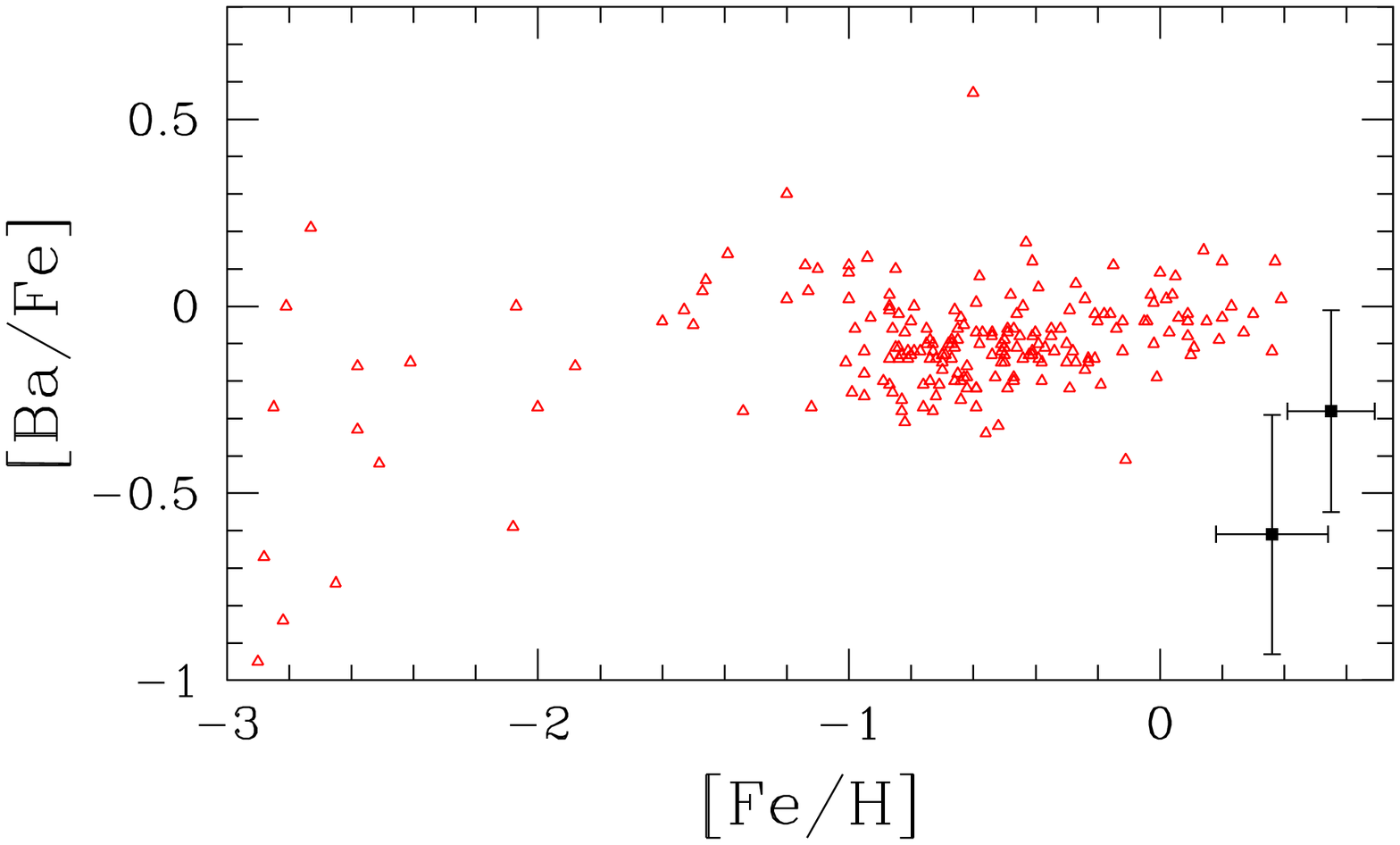}
\caption{[Ba/Fe] for \moaspaces and \oglespace (filled black squares) compared
to field stars (open red triangles). [Ba/Fe] shows the largest deviation from
the trends seen in the disk stars of any element studied in this paper. This
low Ba value is consistent with the idea that the r-process is the dominate
producer of heavy elements in the Bulge. It also puts constraints on 
s-process contributions to Ba (and the accompanying C and N contributions) from
AGB stars. }
\label{fig:ba}
\end{figure}

The [Ba/Fe] for \moaspaces falls considerably below 
the solar value ([Ba/Fe]=$-0.61$ 
(Fig.~\ref{fig:ba}). It falls in the range not seen in 
other parts of Galaxy, except for the metal-poor halo.
We note that we could only measure 1 line of Ba in \moas, and
the solar value we measure is the most discrepant from the solar value
in \citet{grevsauv}. However, even if the Grevesse \& Sauval value is
used, the [Ba/Fe] for \moaspaces is still subsolar ([Ba/Fe]=$-$0.24).
If we use the solar value of Ba (log($\epsilon$(Ba)=2.39) 
reported in \citet{johnson:07}, \oglespace
has a [Ba/Fe]$=-0.28$.

The 
paucity of Ba in the halo stars 
is explained by the fact that the r-process is the only available
channel for producing the heavy elements 
in the early Univere, and the r-process is not an efficient
producer of Ba.
It is tempting to ascribe the
low Ba in \moaspaces to the same cause, especially in light of the
high [Eu/Fe] measurements in Bulge giants by \citet{mcwilliam:94}. 
Stellar populations dominated by Type II SNe and r-process production should
have high [Eu/Fe] and [Eu/Ba] before Type Ia SNe and AGB stars eventually
add Fe and Ba to the ISM. Whether the Ba deficiency in \moaspaces can
be explained by a lack of contributions from AGB stars depends on the
relative timescales of Type Ia SN pollution and AGB pollution and whether
the low [$\alpha$/Fe] values in \moaspaces are the result of Type Ia pollution.
If we assume that Type Ia SNe have contributed significantly to the
abundances of \moas, which is reasonable, then AGB pollution must
trail Type Ia SN production.
The evidence on this point is mixed.
\citet{simmerer:04} saw, in addition to a wide range at
any given metallicity, a rise in the [La/Fe] ratios at
[Fe/H] $>-2$. Because La, like Ba, is mostly due to the s-process, this 
would indicates the s-process from AGB stars is 
added before Fe from Type Ia SNe causes the [$\alpha$/Fe]
ratios to turn over. On the other hand, \citet{melendez:07} 
argued based on the isotope ratios of Mg that 3-6 M$_{\odot}$ stars did
not start contributing to the halo until [Fe/H]$\geq 1.5$, at the
same time or later than the Type Ia SNe.
The [Ba/Fe] measured in \oglespace is closer to the solar value,
and suggests that by [Fe/H]=0.5 the Bulge had reached the same point
in chemical evolution as the solar neighborhood, with substantial
amounts of Ba supplied by the s-process in AGB stars balancing the
iron supplied by Type Ia SNe. 
Ba abundances for Bulge stars with a wide range of metallicity would 
help clarify the origin of the Ba abundance in the Bulge.

\section{Conclusions}

The results for \moaspaces demonstrate the unique information that can
be obtained from high-resolution spectra of microlensed dwarfs in 
the Bulge. Because microlensing is not biased in metallicity, we
can measure the MDF of the Bulge main-sequence stars when sufficient
number of highly magnified Bulge dwarfs have been observed with high-dispersion
spectrographs. However,
the two dwarfs we have studied so far both have [Fe/H]$> 0.30$, which
makes them more metal-rich than any of the M giants observed at 
high-resolution. If the temperatures and metallicities measured for
the two dwarfs are accurate, then these stars
are younger than the bulk of the Bulge population.

For many elements, the abundance ratios we measure in \moaspaces are not
surprising. The iron-peak and $\alpha$-elements (O, Mg, Si, Ca, and Ti) fall
on the trend observed in the Bulge giants. The solar values for the
 [$\alpha$/Fe] ratios suggest that Type Ia SN ejecta are responsible
for some of the Fe, which would be expected if these metal-rich stars 
were part of a
younger population in the Bulge. The [C/Fe] and [N/Fe] ratios
are also $\sim$ 0, suggesting that AGB stars have also begun to contribute
these light elements to offset the contribution of Fe from Type Ia SNe.
It is therefore unexpected to find [Ba/Fe]$\sim$-$0.5$. Such low
[Ba/Fe] values, when found in the halo, imply that only Type II SNe
and the r-process have polluted the gas, and not the s-process from AGB stars.

The C and N abundances we measure in \moaspaces are unaffected by any mixing
or dilution and therefore are the result of the chemical evolution of
the Bulge. Comparing the total
amount of C+N to that measured in Bulge giants, we conclude that only
mild mixing, as expected in theoretical models, has occurred in the
Bulge. This agrees with the assertions by 
\citet{lecureur:07}, \citet{zoccali:06}, and \citet{mcwilliam:07} that the O, Na and Al abundances in
Bulge giants should not have been affected by mixing. 
The [Na/Fe] and [Al/Fe] ratios for \moaspaces fall on the
lower end of the values seen in Bulge giants of this metallicity. The
[Na/Fe] and [Al/Fe] ratios in giants are marked above all else by
large scatter and more abundance ratios in dwarfs are needed before a
comparison of scatter can be made.

We can use our measurements of the abundances of K, Cu and Zn, rare for Bulge
stars, to exploit the different star formation history of the Bulge
to study the nucleosynthesis sites of these elements. The supersolar [Cu/Fe] and
subsolar [Zn/Fe] values agree with the model of \citet{bisterzo:05} and the 
production of these two elements predominantly in Type II SNe. The
chemical evolution of K is more difficult to understand, because the 
[K/Mg] value in the Bulge is similar to that in the metal-poor halo, although
the production of K should be enhanced in metal-rich SNe.

Finally, we note that the standard microlensing technique of estimating 
the intrinsic color and magnitude of the source star by using the
offset from the position of the red clump can be tested by deriving
a temperature and gravity from the spectrum. For \moas, the photometric
and spectroscopic temperature agree very well, although the agreement
was considerably less good for \ogle.

In summary, the abundances in \moaspaces provide unique information about
the formation and evolution of the Bulge. Larger samples of dwarfs observed
in this way would allow the derivation of the dwarf MDF and the trends in abundance ratios 
over a wide range of metallicity. By taking advantage of microlensing events
in the Bulge, we can achieve this goal with a modest amount of telescope time.

\acknowledgments
Our thanks to Ian Thompson for help with the echelle observations at
Las Campanas. Thomas Masseron and Bertrand Plez provided invaluable
support in installing and running TurboSpectrum. We would also like to
thank Daniel Kelson for codes and
support for reducing the MIKE data. Jon Fulbright, Manuela
Zoccali, and Solange Rami\'rez kindly provided metallicity data for the
bulge giant metallicity distribution functions.
We acknowledge support from:
NSF AST-042758 and NASA NNG04GL51G (AG). Work by B.S.G. was partially supported by a Menzel Fellowship
from the Harvard College Observatory.


\begin{deluxetable}{lrrrrrr}
\tablenum{1}\label{Tab:EW}
\tablewidth{0pt}
\tablecaption{Line Parameters and Equivalent Widths}
\tablehead{
\colhead{Ion} & \colhead{Wavelength} & \colhead{E.P.} & \colhead{log \it{gf}}
& \colhead{EW$_{star}$} & \colhead{EW$_{sun}$} & \colhead{Source} 
}
\startdata
\ion{O}{1} & 7771.941 & 9.15&  0.370  &101.5     &70.2 & 1\\  
\ion{O}{1} & 7774.161 & 9.15&  0.220  & 86.7     & 60.1& 1\\  
\ion{O}{1} & 7775.390 & 9.15&  0.000  & 71.1     & 47.4& 1\\  
\ion{Na}{1}& 5682.633 & 2.10& $-$0.710 & 148.1   & 96.9& 2\\  
\ion{Na}{1}& 5688.194 & 2.10& $-$0.400 & 150.6   &118.7 &2 \\  
\ion{Na}{1}& 6154.226 & 2.10& $-$1.570  & 60.9   & 35.9& 2\\  
\ion{Na}{1}& 6160.747 & 2.10& $-$1.270  & 82.3   & 55.8& 2\\  
\ion{Mg}{1}&  5711.088&  4.33& $-$1.870 & 140.6  &112.7 &2 \\  
\ion{Mg}{1}&  7387.689&  5.75& $-$1.270 & 141.2 & 70.6& 2\\  
\ion{Al}{1} & 7836.134&  4.02& $-$0.650 &  93.1 & 54.2& 2\\  
\ion{Si}{1}&  5665.555&  4.93& $-$1.940 &  89.2 & 40.2& 2\\  
\ion{Si}{1}&  5690.425&  4.93& $-$1.770 &  57.4 & 49.7& 2\\  
\ion{Si}{1}&  5772.146&  5.08& $-$1.650 &  69.9 & 51.9& 2\\  
\ion{Si}{1}&  6125.021&  5.61& $-$1.520 &  57.7 & 30.4& 2\\  
\ion{Si}{1}&  6142.483&  5.62& $-$1.500 &  61.8 & 32.3& 2\\  
\ion{Si}{1}&  6155.134&  5.62& $-$0.720 & 124.5 & 84.5& 2\\  
\ion{Si}{1}&  6237.319&  5.61& $-$1.050 &  90.9 & 63.9& 2\\  
\ion{Si}{1}&  6244.466&  5.62& $-$1.320 &  77.6 & 47.3& 2\\  
\ion{Si}{1}&  7415.948&  5.61& $-$0.850 & 115.0 & 85.5& 2\\  
\ion{Si}{1}&  7918.384&  5.95& $-$0.510 & 157.5 & 84.5& 2\\  
\ion{Si}{1}&  7932.348&  5.96& $-$0.370 & 131.8 & 92.3& 2\\  
\ion{Si}{1}&  7944.001&  5.98& $-$0.210 & 140.8 & 107.0& 2\\  
\ion{K}{1}  & 7698.974 & 0.00& $-$0.170&   syn  &  syn& 1\\  
\ion{Ca}{1}&  5581.965&  2.52& $-$0.530 & 111.8 & 94.7& 3\\  
\ion{Ca}{1}&  5590.114&  2.52& $-$0.571 & 148.4 & 91.0& 3\\  
\ion{Ca}{1}&  6161.297&  2.52& $-$1.266 &  77.3 & 59.2& 3\\  
\ion{Ca}{1}&  6166.439&  2.52& $-$1.142 &  83.2 & 70.2& 3\\  
\ion{Ca}{1}&  6169.042&  2.52& $-$0.797 & 110.2 & 88.7& 3\\  
\ion{Ca}{1}&  6169.563&  2.52& $-$0.478 & 147.8 & 105.2&3 \\  
\ion{Ca}{1}&  6455.598 & 2.52& $-$1.290 &  59.6 & 56.9& 3\\  
\ion{Ca}{1}&  6471.662&  2.52& $-$0.686 & 116.1 & 90.2& 3\\  
\ion{Ca}{1}&  6499.650&  2.52& $-$0.818 &  97.5 & 84.1& 3\\  
\ion{Sc}{2} & 5526.820 & 1.77& 0.02    &   syn  &  syn& 5\\  
\ion{Sc}{2} & 5657.880 & 1.51& $-$0.600 &   syn  &  syn& 5\\  
\ion{Sc}{2} & 6245.620 & 1.51& $-$1.070 &   syn  &  syn& 5\\  
\ion{Ti}{1}&  5866.451&  1.07& $-$0.780 &  60.6 & 48.1& 2\\  
\ion{Ti}{1}&  6126.216&  1.07& $-$1.370 &  59.7 & 23.4& 2\\  
\ion{Ti}{1}&  6258.102&  1.44& $-$0.300 &  72.9 & 49.8& 2\\  
\ion{Ti}{1}&  6261.098&  1.43& $-$0.420 &  61.1 & 50.1& 2\\  
\ion{Ti}{2}&  5381.015&  1.57& $-$1.920 &  80.4 & 59.3& 4\\  
\ion{Mn}{1} & 5470.270 & 2.14& $-$1.460 &   syn  &  syn& 5\\  
\ion{Mn}{1} & 6013.520 & 3.07& $-$0.250 &   syn  &  syn& 5\\  
\ion{Mn}{1} & 6021.710 & 3.07& 0.03    &   syn  &  syn& 5\\  
\ion{Fe}{1}&  5307.361&  1.61& $-$2.987 & 105.7 & 84.5& 1\\  
\ion{Fe}{1} & 5321.108&  4.43& $-$0.951 &  65.1 & 40.8& 1\\  
\ion{Fe}{1}&  5322.041&  2.28& $-$2.803 & 105.4 & 59.8& 1\\  
\ion{Fe}{1} & 5329.989&  4.08& $-$1.189 & 128.7 & 55.7& 1\\  
\ion{Fe}{1}&  5379.574&  3.70& $-$1.510 &  74.2 &  60.5&1 \\  
\ion{Fe}{1}&  5501.465&  0.96& $-$3.047 & 127.1 & 115.4&1 \\  
\ion{Fe}{1}&  5506.779&  0.99& $-$2.797 & 149.4 & 115.1&1 \\  
\ion{Fe}{1} & 5701.545&  2.56& $-$2.120 & 114.7 & 81.9& 1\\  
\ion{Fe}{1}&  5705.465&  4.30& $-$1.500 &  54.3 & 38.7& 1\\  
\ion{Fe}{1}&  6027.051&  4.08& $-$1.090 &  78.4 & 65.1& 1\\  
\ion{Fe}{1}&  6082.711&  2.22& $-$3.573 &  63.5 & 34.0& 1\\  
\ion{Fe}{1}&  6165.360 & 4.14& $-$1.470 &  72.3 & 43.7& 1\\  
\ion{Fe}{1}&  6173.336 & 2.22& $-$2.880 & 119.9 & 68.7& 1\\  
\ion{Fe}{1}&  6219.281 & 2.20& $-$2.420 & 114.4 & 86.6& 1\\  
\ion{Fe}{1}&  6252.555 & 2.40& $-$1.690 & 148.8 & 115.3&1 \\  
\ion{Fe}{1}&  6265.134 & 2.18& $-$2.550 & 107.8 & 87.2& 1\\  
\ion{Fe}{1}&  6297.793 & 2.22& $-$2.740&   94.6 & 72.8& 1\\  
\ion{Fe}{1}&  6301.501 & 3.65& $-$0.718&  141.5 & 121.6&1 \\  
\ion{Fe}{1}&  6311.500 & 2.83& $-$3.141&   49.0 & 27.1& 1\\  
\ion{Fe}{1}&  6322.685 & 2.59& $-$2.426&  100.1 & 79.0& 1\\  
\ion{Fe}{1}&  6335.331 & 2.20& $-$2.180&  127.5 & 95.8& 1\\  
\ion{Fe}{1}&  6344.149 & 2.43& $-$2.920&   91.7 & 58.4& 1\\  
\ion{Fe}{1}&  6380.743 & 4.19& $-$1.380&   82.6 & 51.6& 1\\  
\ion{Fe}{1}&  6408.018 & 3.69& $-$1.018&  131.7 & 94.7& 1\\  
\ion{Fe}{1}&  6411.649 & 3.64& $-$0.720&  139.5 & 128.7&1 \\  
\ion{Fe}{1}&  6498.939 & 0.96& $-$4.700&   49.8 & 45.0& 1\\  
\ion{Fe}{1}&  6592.914 & 2.73& $-$1.470&  129.6 & 107.3&1 \\  
\ion{Fe}{1}&  6593.870 & 2.43& $-$2.420&  117.3 & 81.7& 1\\  
\ion{Fe}{1}&  6804.001 & 4.65& $-$1.496&   67.2 & 21.1& 1\\  
\ion{Fe}{1}&  6810.263 & 4.61& $-$0.986&   71.4 & 48.9& 1\\  
\ion{Fe}{1}&  6855.162 & 4.56& $-$0.740&  102.2 & 69.0& 1\\  
\ion{Fe}{1}&  6858.150 & 4.61& $-$0.930&   79.5 & 50.8& 1\\  
\ion{Fe}{1} & 7401.685 & 4.19& $-$1.350&   58.7 & 39.7& 1\\  
\ion{Fe}{1} & 7710.364 & 4.22& $-$1.110&  103.9 & 63.5& 1\\  
\ion{Fe}{1} & 7941.089 & 3.27& $-$2.580&   59.8 & 42.0& 1\\  
\ion{Fe}{2} & 6247.557 & 3.89& $-$2.430&   86.3 & 51.8& 2\\  
\ion{Fe}{2}  &6416.919 & 3.89& $-$2.880&   65.0 & 41.0& 2\\  
\ion{Fe}{2} & 6432.680 & 2.89& $-$3.690&   61.3 & 40.0& 2\\  
\ion{Fe}{2} & 6456.383 & 3.90& $-$2.190&   71.7 & 61.2& 2\\  
\ion{Ni}{1} & 5754.655 & 1.93& $-$1.850&  133.5 & 75.4& 2\\  
\ion{Ni}{1} & 5805.213 & 4.17& $-$0.620&   60.5 & 42.1& 2\\  
\ion{Ni}{1} & 6108.107 & 1.68& $-$2.430&   68.5 & 63.3& 2\\  
\ion{Ni}{1} & 6111.066 & 4.09& $-$0.820&   70.0 & 32.7& 2\\  
\ion{Ni}{1} & 6128.963 & 1.68& $-$3.360&   46.8 & 24.8& 2\\  
\ion{Ni}{1} & 6175.360 & 4.09& $-$0.500&   63.5 & 46.9& 2\\  
\ion{Ni}{1} & 6176.807 & 4.09& $-$0.260&   95.1 & 63.4& 2\\  
\ion{Ni}{1} & 6314.653 & 1.94& $-$2.000&   92.1 & 72.4& 2\\  
\ion{Ni}{1} & 6378.247 & 4.15& $-$0.810&   59.5 & 29.9& 2\\  
\ion{Ni}{1} & 6482.796 & 1.93& $-$2.760&   54.8 & 40.7& 2\\  
\ion{Ni}{1} & 6598.593 & 4.24& $-$0.910&   42.0 & 24.8& 2\\  
\ion{Ni}{1} & 6643.629 & 1.68& $-$1.910 & 122.2 & 95.0& 2\\  
\ion{Ni}{1} & 6767.768 & 1.83& $-$2.100&  106.2 & 80.5& 2\\  
\ion{Ni}{1} & 6772.313 & 3.66& $-$0.940&   83.8 & 49.8& 2\\  
\ion{Ni}{1} & 7110.892 & 1.94& $-$2.880&   57.8 & 36.0& 2\\  
\ion{Ni}{1} & 7715.583 & 3.70& $-$1.010&  105.7 & 49.7& 2\\  
\ion{Ni}{1} & 7748.891 & 3.71& $-$0.330&  164.4  & 84.0&2 \\  
\ion{Ni}{1} & 7788.936 & 1.95& $-$1.750&  123.4 & 92.2& 2\\  
\ion{Ni}{1} & 7797.586 & 3.90& $-$0.320&  103.6 & 79.2& 2\\
\ion{Cu}{1} & 5782.050 & 1.64& $-$2.920 &   syn  &  syn& 5\\  
\ion{Zn}{1} & 4722.153 & 4.03& $-$0.338&   syn  &  syn& 1\\  
\ion{Zn}{1} & 4810.528 & 4.08& $-$0.137&   syn  &  syn& 1\\  
\ion{Ba}{2} & 6496.410 & 0.60& $-$0.380&   syn  &  syn& 5\\  
\enddata
\tablerefs{(1) VALD database, \citet{vald}, (2) \citet{bensby:03}, (3) \citet{smith:ca},
(4) \citet{pickering:ti}, (5)\citet{johnson:06} and references therein}
\end{deluxetable}

\begin{deluxetable}{lrrrrrrrr}
\tablenum{2}\label{Tab:Abund}
\tablewidth{0pt}
\tablecaption{Abundances}
\tablehead{
\colhead{Ion} & \colhead{log$\epsilon$} & \colhead{$\sigma_{\epsilon}$}
& \colhead{[X/FeI]\tablenotemark{a}} & \colhead{$\sigma_{[X/FeI]}$} & \colhead{$\sigma$}
& \colhead{N$_{\rm lines}$} & \multicolumn{2}{c}{Solar} \\
\colhead{} & \colhead{} & \colhead{}
& \colhead{} & \colhead{} & \colhead{}
& \colhead{} & \colhead{meas.} & \colhead{GS98} 
}
\startdata
C (CH) &     8.8  & 0.26& 0.04& 0.22& 0.20& \ldots & 8.4  & 8.52\\
N (CN) &     8.2  & 0.56& $-$0.06& 0.43& 0.20& \ldots & 7.9  & 7.92\\
\ion{O}{1}  &9.09 & 0.11& $-$0.16& 0.22& 0.33& 3      & 8.89 & 8.83\\
\ion{Na}{1} &6.69 & 0.13& 0.09& 0.16& 0.12& 4      & 6.24 & 6.33\\
\ion{Mg}{1} &8.18 & 0.21& 0.16& 0.25& 0.17& 2      & 7.66 & 7.58\\
\ion{Al}{1} &6.98 & 0.16& 0.11& 0.22& 0.10& 1      & 6.51 & 6.47\\
\ion{Si}{1} &7.98 & 0.10& 0.06& 0.13& 0.10& 12     & 7.56 & 7.55\\
\ion{K}{1}  &5.60 & 0.28& 0.14& 0.29& 0.20& 1      & 5.10 & 5.12\\
\ion{Ca}{1} &6.58 & 0.29& $-$0.09& 0.16& 0.23& 9     & 6.31 & 6.36\\
\ion{Sc}{2} &3.25 & 0.31& $-$0.30& 0.23& 0.13& 3      & 3.19 & 3.17\\
\ion{Ti}{1} &5.25 & 0.23& $-$0.05& 0.17& 0.27& 4      & 4.94 & 5.02\\
\ion{Ti}{2} &5.18 & 0.36& $-$0.22& 0.28& 0.20& 1      & 5.04 & 5.02\\
\ion{Mn}{1} &5.62 & 0.13& $-$0.09& 0.12& 0.08& 3      & 5.34 & 5.39\\
\ion{Fe}{1} &7.81 & 0.18& 0.36& \ldots&0.25 & 35     & 7.45 & 7.50\\
\ion{Fe}{2} &7.81 & 0.28& $-$0.18& 0.19& 0.25& 4      & 7.63 & 7.50\\
\ion{Ni}{1} &6.61 & 0.10& 0.02& 0.09& 0.26& 19     & 6.23 & 6.25\\
\ion{Cu}{1} &4.40 & 0.23& $-$0.01& 0.20& 0.10& 1      & 4.05 & 4.21\\
\ion{Zn}{1} &4.63 & 0.23& $-$0.28& 0.21& 0.10& 2      & 4.55 & 4.60\\
\ion{Ba}{2} &2.25 & 0.43& $-$0.61& 0.32& 0.20& 1      & 2.50 & 2.13\\
\enddata
\tablenotetext{a}{[X/\ion{Fe}{1}] given for all species except \ion{Fe}{1},
where [\ion{Fe}{1}/H] is given.}
\end{deluxetable}

\begin{deluxetable}{lcccccccccccccccc}
\rotate
\tabletypesize{\footnotesize}
\tablenum{3}\label{Tab:Lit}
\tablecaption{Literature Sources}
\tablehead{
\colhead{Source} &
\colhead{C} & \colhead{N} & \colhead{O} & \colhead{Na} &
\colhead{Mg} & \colhead{Al} & \colhead{Si} & \colhead{K} &
\colhead{Ca} & \colhead{Sc} & \colhead{Ti} & \colhead{Mn} & \colhead{Ni} & \colhead{Cu} & \colhead{Zn} &
\colhead{Ba} 
}
\startdata
\multicolumn{17}{c}{Bulge Stars} \\
\citet{fulbright:07} &   &   & x & x & x & x & x &   & x &   &   &   &   &   &   &  \\
\citet{rich:05}      &   &   &   &   &   &   &   &   &   &   &   &   &   &   &   &  \\
\citet{rich:07}      & x &   & x &   & x &   & x &   & x &   & x &   &   &   &   &  \\
\citet{lecureur:07}  &   &   &   &   &   &   &   &   &   &   &   &   &   &   &   &  \\
\citet{cunha:06}     & x & x & x & x &   &   &   &   &   &   & x &   &   &   &   &  \\
\multicolumn{17}{c}{Halo/Disk Data} \\
\citet{reddy:06}     & x &   & x & x & x & x & x &   & x & x & x & x & x & x & x & x\\
\citet{reddy:03}     & x & x & x & x & x & x & x &   & x & x & x & x & x & x & x & x\\
\citet{feltzing:07}  &   &   &   &   &   &   &   &   &   &   &   & x &   &   &   &  \\
\citet{bensby:06}    & x &   &   &   &   &   &   &   &   &   &   &   &   &   &   &  \\
\citet{bensby:05}    &   &   &   &   & x &   & x &   &   &   & x &   & x &   & x & x\\
\citet{bensby:04}    &   &   & x &   &   &   &   &   &   &   &   &   &   &   &   &  \\
\citet{bensby:03}    &   &   &   & x & x &   &   &   &   &   &   &   &   &   &   &  \\
\citet{chen:04}      &   &   &   & x & x & x & x &   & x & x & x & x & x &   &   & x\\
\citet{carretta:00}  & x & x &   &   &   &   &   &   &   &   &   &   &   &   &   &  \\
\citet{zhang:06}     &   &   &   &   &   &   &   & x &   &   &   &   &   &   &   &  \\
\enddata
\end{deluxetable}

\end{document}